\documentstyle[amsfonts,epsfig,here,portland]{aipproc}

\def\beq{\begin{equation}}
\def\eeq{\end{equation}}
\def\bea{\begin{eqnarray}}
\def\eea{\end{eqnarray}}
\def\bq{\begin{quote}}
\def\eq{\end{quote}}

\def\PR{{\it Phys.Rev.} }

\def\gappeq{\mathrel{\rlap {\raise.5ex\hbox{$>$}}
{\lower.5ex\hbox{$\sim$}}}}

\def\lappeq{\mathrel{\rlap{\raise.5ex\hbox{$<$}}
{\lower.5ex\hbox{$\sim$}}}}

\begin{document}

\begin{flushright}
CERN-TH/99-358 \\
astro-ph/9911440
\end{flushright}
\title{Particles and Cosmology: Learning from Cosmic Rays}

\author{John Ellis$^*$ }
\address{$^*$Theoretical Physics Division, CERN \\ CH -- 1211 Geneva 23}
\maketitle

\begin{center}
{\it Contribution to the Proceedings of the 26th International
Cosmic-Ray Conference, Salt Lake City, August 1999}
\end{center}

\begin{abstract}
The density budget of the Universe is reviewed, and then specific particle
candidates for non-bayonic dark matter are introduced, with emphasis on
the
relevance of cosmic-ray physics. The sizes of the {\it neutrino masses}
indicated by recent atmospheric and solar neutrino experiments may be too
small to contribute much hot dark matter.  My favoured candidate for the
dominant cold dark matter is the {\it lightest supersymmetric particle},
which probably weighs between about 50~GeV and about 600~GeV. Strategies
to search for it via cosmic rays due to annihilations in the halo, Sun and
Earth, or via direct scattering experiments, are mentioned. Possible {\it
superheavy relic particles} are also discussed, in particular metastable
string- or $M$-theory {\it cryptons}, that may be responsible for the
ultra-high-energy cosmic rays. Finally, it is speculated that a non-zero
contribution to the {\it cosmological vacuum energy} might result from
incomplete relaxation of the quantum-gravitational vacuum. 
\end{abstract}

\section{Density Budget of the Universe}
 
As you know, the Universe becomes almost homogeneous and
isotropic, when viewed on a sufficiently large scale. This
suggests very strongly that it may be described approximately by
a Robertson-Walker-Friedmann metric. The crucial parameters
describing the expansion of the Universe are then its density
$\rho$ and the curvature of the Universe; $ k = 0, +1$ or $-1$
for a critical, closed or open Universe, respectively. If $k = 0$,
the density must equal the critical density $\rho_c \sim 2 h^2 \times
10^{-29}$ g/cm$^3$, where $h$ is the present Hubble expansion rate
in units of 100 km/s/Mpc.
Much of the subsequent cosmological discussion is phrased in terms of  the
density budget of
the Universe, expressed as contributions relative to the critical density:
$\Omega_i \equiv \rho_i/\rho_c$.

$\Omega_{tot}$: Inflation suggests that this is practically indistinguishable
from unity: 
$\Omega_{tot} = 1 \pm {\cal O}(10^{-4})$ \cite{Olive}, although there are some models that predict
$\Omega_{tot} < 1$~\cite{openinflation}. However, the data on the small
anisotropies in the cosmic Microwave Background (CMB)~\cite{anisotropies}
support the inflationary suggestion that $\Omega_{tot} \sim 1$ and
$\Omega_k \sim 0$, as 
summarized in Fig.~1~\cite{triangle}.

\begin{figure}[htb] 
\centerline{\epsfig{file=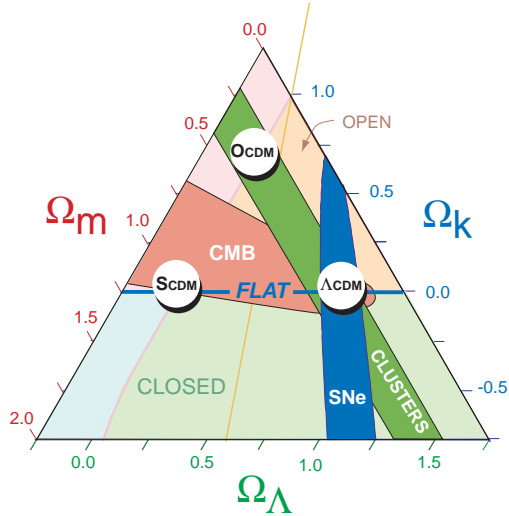,width=7cm}}
\vspace{10pt}
\hspace{3pt}
\caption[]{Compilation~\cite{triangle} of constraints on contributions to
the cosmological energy density, as provided by the cosmic microwave
background (CMB), cluster data and high-redshift supernovae.
Concordance appears for the model $\Lambda_{CDM}$ with
cosmological vacuum energy, but not for standard cold dark
matter $S_{CDM}$ or for an open dark matter model $O_{CDM}$.}
\label{fig1}
\end{figure}

$\Omega_{b}$: Measurements of the $D/H$ ratio in high-redshift Lyman-$\alpha$
clouds~\cite{Tytler} correspond to
\beq
{D\over H} = (3.3 \pm 0.3)\times 10^{-5}
\label{one}
\eeq
If this is indeed the correct primordial $D/H$ ratio, Big-Bang nucleosynthesis
calculations suggest that~\cite{Tytler}
\beq
{n_B\over s} = (5.1 \pm 0.3) \times 10^{-10}
\label{two}
\eeq
corresponding to
\beq
\Omega_B h^2 = 0.019 \pm 0.001
\label{three}
\eeq
Using the currently favoured range $h = 0.65 \pm 0.10$, we
see from
(\ref{twenty}) that $\Omega_b \lappeq 0.08$, which is insufficient to explain
all the matter density in the following paragraph.

$\Omega_m$: The cluster measurements ($M/L$ ratio, present and past abundances,
cluster dynamics and the baryon fractions inferred from $X$-ray measurements)
all suggest~\cite{Bahcall}
\beq
\Omega_m \sim 0.2~~{\rm to}~~ 0.3
\label{four}
\eeq
as also seen in Fig. 1. Moreover,  the combination of CMB measurements and
high-redshift supernovae~\cite{highz} also support 
independently such a value for $\Omega_m$.

$\Omega_{CDM}$: The theory~\cite{triangle} of large-scale structure
formation strongly suggests
that most of $\Omega_m$ is cold dark matter, so that
\beq
\Omega_{CDM} \sim \Omega_m
\label{five}
\eeq
as perhaps provided by the supersymmetric particles discussed later.

$\Omega_{HDM}$:
The theory~\cite{triangle} of structure formation also suggests that 
the density of hot dark matter $\Omega_{HDM} \ll
\Omega_{CDM}$.
The present and prospective sensitivities of cosmological data to $m_\nu$ are
shown in Fig. 2~\cite{Tegmark}. So far, $m_\nu \gappeq$ 3 eV is excluded
\begin{figure}[htb] 
\centerline{\epsfig{file=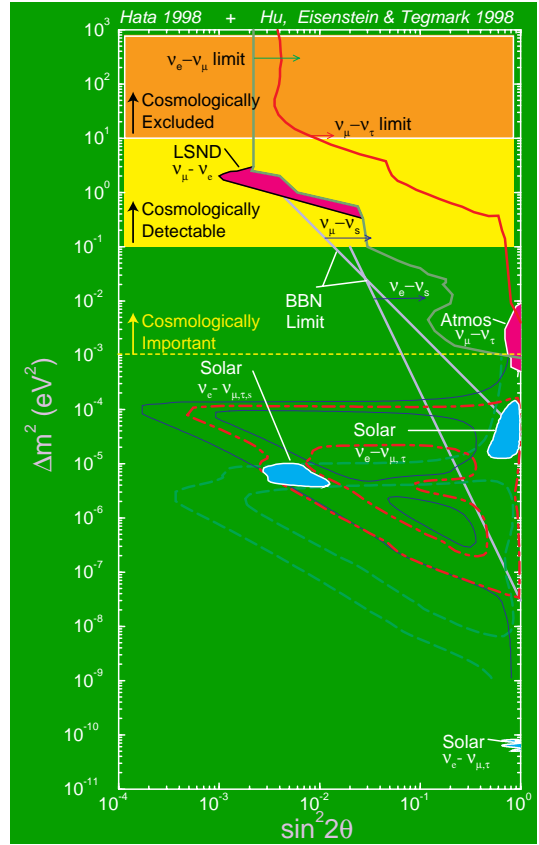,width=7cm}}
\vspace{10pt}
\hspace{3pt}
\caption[]{Compilation~\cite{Tegmark} of indications on neutrino
mass-squared
differences $
 \Delta m^2$ and mixing angles $\theta$ from oscillation experiments, compared
 with cosmological sensitivities to neutrino masses.}
\label{fig2}
\end{figure}
by the available
upper limit on the density of hot dark matter,
whereas the possible comparison of future data on
large-scale structure and the CMB are thought to be sensitive to $m_\nu
\gappeq$ 0.3 eV. This is somewhat above the range $m_\nu \sim$ 0.1 to
0.03 eV
favoured by the atmospheric neutrino data, but one should not abandon hope
of
detecting neutrino masses astrophysically~\cite{Tegmark}.
As discussed in Section 2, the indications of neutrino masses
from
atmospheric and solar neutrino data can most easily be explained by
light neutrinos: $m_{\nu_i} <$ 0.1 eV, which would make only a  small
contribution to $\Omega_{tot}$.

$\Omega_\Lambda$: If one follows the inflationary path
supported by the CMB~\cite{anisotropies}, so that
$\Omega_{tot} \sim 1$, and takes at face value the suggestions from
cluster measurements that $\Omega_m \sim 0.3$, then
the largest fraction of the energy
density of the Universe may be provided by vacuum energy:
$\Omega_{\Lambda} \sim 0.7$.
\begin{figure}[htb] 
\centerline{\epsfig{file=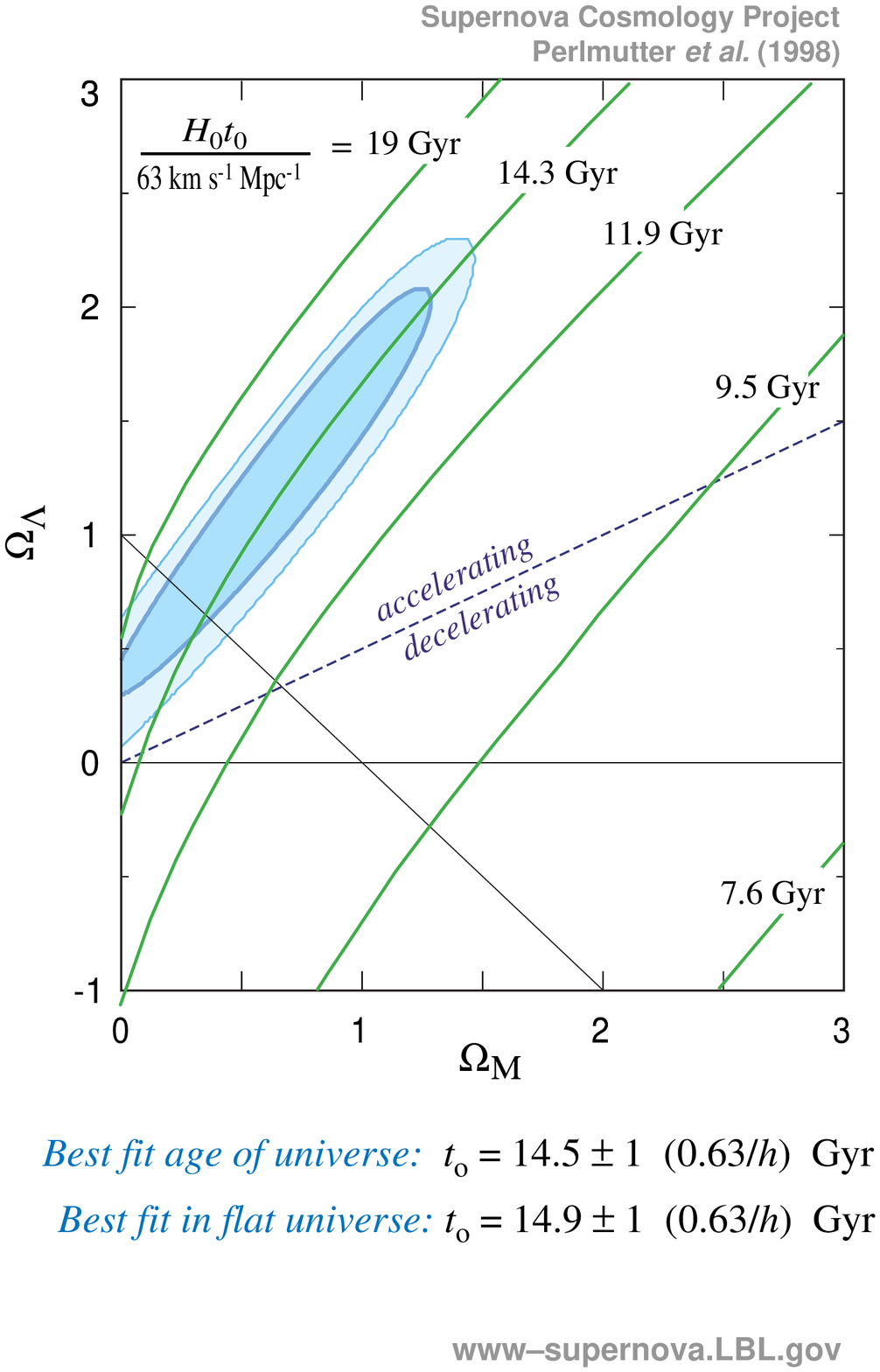,width=8cm}}
\vspace{10pt}
\hspace{3pt}
\caption[]{Constraint on $\Omega_\Lambda, \Omega_M$ from one
set of high-redshift supernova data~\cite{highz}.}
\label{fig3}
\end{figure}
This scenario is supported by the recent high-redshift supernova
data~\cite{highz} shown in Fig.~3, which suggest that $\Omega_{\Lambda} - 
\Omega_m \sim 0.4$. Combining this estimate with the suggestion of
inflation that $\Omega_{tot} = \Omega_m + \Omega_\Lambda \simeq 1$,
one recovers independently the preference for
$\Omega_m \sim 0.3, \Omega_{\Lambda} \sim 0.7$.

A remarkably consistent picture of the density budget of the Universe may be
emerging:
\beq
\Omega_{tot} \simeq 1 = \Omega_m + \Omega_\Lambda : \\
\Omega_m \sim 0.3, \Omega_\Lambda \sim 0.7
\label{six}
\eeq
where
\beq
\Omega_m = \Omega_{CDM} + \Omega_\nu + \Omega_b
\label{seven}
\eeq
with
\beq
\Omega_b < 0.1~,~~~\Omega_\nu \ll \Omega_{CDM} \simeq \Omega_m
\label{eight}
\eeq
It remains to be seen whether future data confirm this picture. For the moment, let us
examine particle candidates for $\Omega_{HDM}$ and $\Omega_{CDM}$, emphasizing their
cosmic-ray manifestations.

\section{Neutrino Masses}

If these are non-zero, laboratory experiments tell us that  they must be much smaller than
those of the corresponding charged leptons~\cite{PDG}:
\beq
m_{\nu_e} \lappeq 2.5~{\rm eV}~, \quad\quad 
m_{\nu_\mu} \lappeq 160~{\rm keV}~, \quad\quad 
m_{\nu_\tau} \lappeq 15~{\rm eV}~,
\label{nine}
\eeq
so one might think naively that they should vanish entirely. However, theorists
believe that particle masses can be strictly zero only 
if there is a corresponding
conserved charge associated with an exact gauge symmetry, which is not the case
for lepton number. Indeed, non-zero neutrino masses appear generically in Grand
Unified Theories (GUTs)~\cite{Peccei}. However, it is not even necessary to
postulate new particles
to get $m_\nu \not= 0$. These could be generated by a non-renormalizable
interaction among Standard Model particles~\cite{BEG}:
\beq
{(\nu_LH)~(\nu_LH)\over M}
\label{ten}
\eeq
where $M \gg m_W$ is some new, heavy mass scale. 
The most plausible guess, though,
is that this heavy mass is that of some heavy particle, perhaps a right-handed
neutrino $\nu_R$ with mass $M \sim M_{GUT}$.

In this case, one expects to find the characteristic see-saw~\cite{seesaw} 
form of neutrino mass matrix:
\beq
(\nu_L , \nu_R)~~\left(\matrix{0&m\cr
m&M}\right)~~\left(\matrix{\nu_L\cr\nu_R}\right)
\label{eleven}
\eeq
where the off-diagonal matrix entries 
in (\ref{eleven}) break SU(2) and have the form of
Dirac mass terms, so that one expects $m = {\cal O}(m_{\ell,q})$. Diagonalizing
(\ref{eleven}), one finds generically a light neutrino mass
\beq
m_\nu \simeq {m^2\over M}
\label{twelve}
\eeq
Choosing representative numbers $m \sim$ 10 GeV, 
$m_\nu \sim 10^{-2}$~eV, one finds
$M \sim 10^{13}$ GeV, in the general ballpark of the grand unification scale.

As you know, data on both solar and atmospheric 
neutrinos favour neutrino oscillations
associated with neutrino mass differences: 
$\nu_e\rightarrow\nu_x$ in the solar case and
$\nu_\mu\rightarrow\nu_x$ in the atmospheric case. (Both of these can be
regarded as
cosmic-ray phenomena!) There are three possible interpretations of the
solar-neutrino data:
vacuum oscillations with $\Delta m^2 \sim 10^{-10}$ eV$^2$ 
and large mixing, and
matter-enhanced MSW oscillations with 
$\Delta m^2 \sim 10^{-5}$ eV$^2$ and either large or
small mixing, as seen in Fig. 4~\cite{bahcallpage}.
\begin{figure}[htb] 
\centerline{\epsfig{file=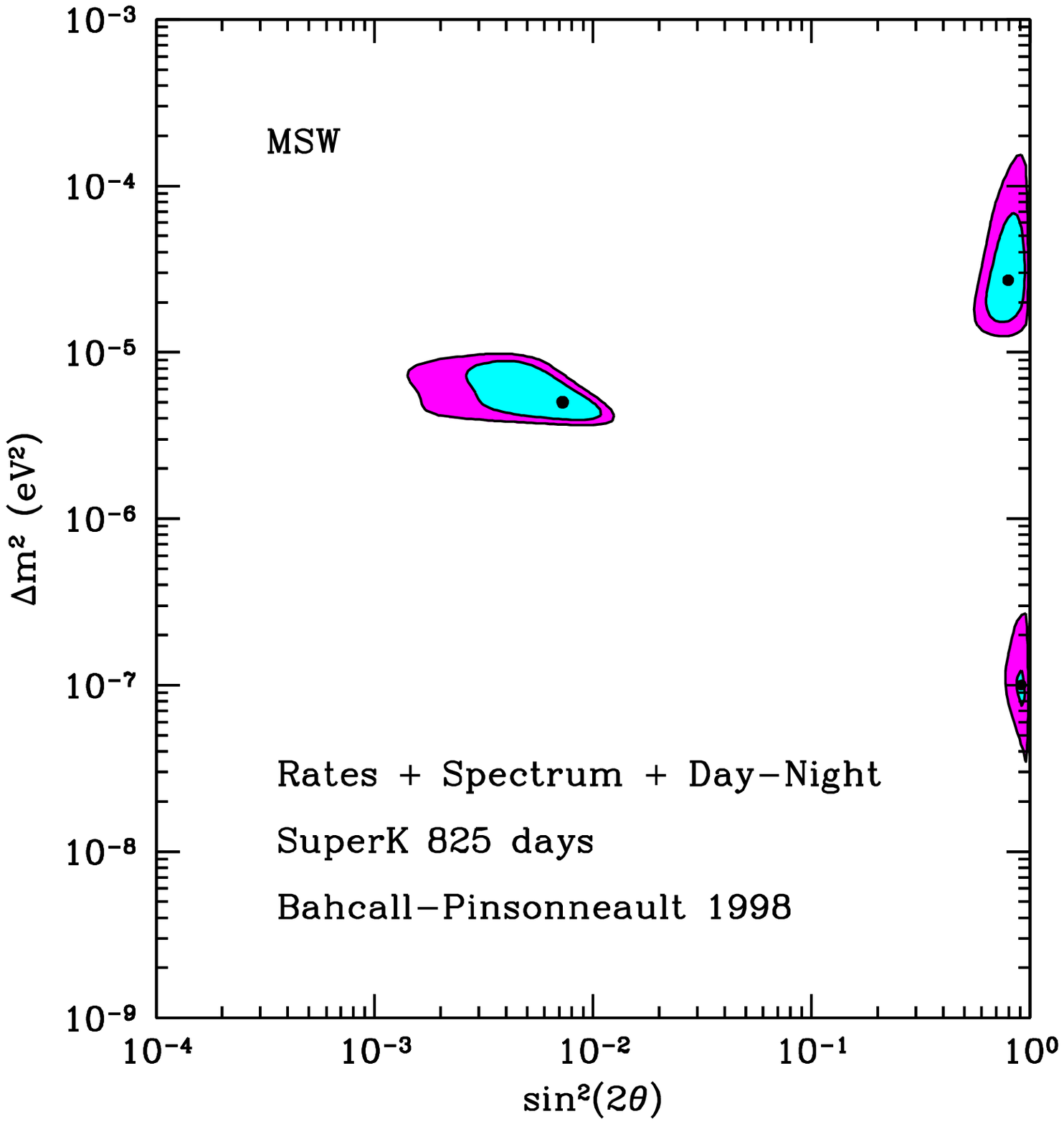,width=7cm}
\epsfig{file=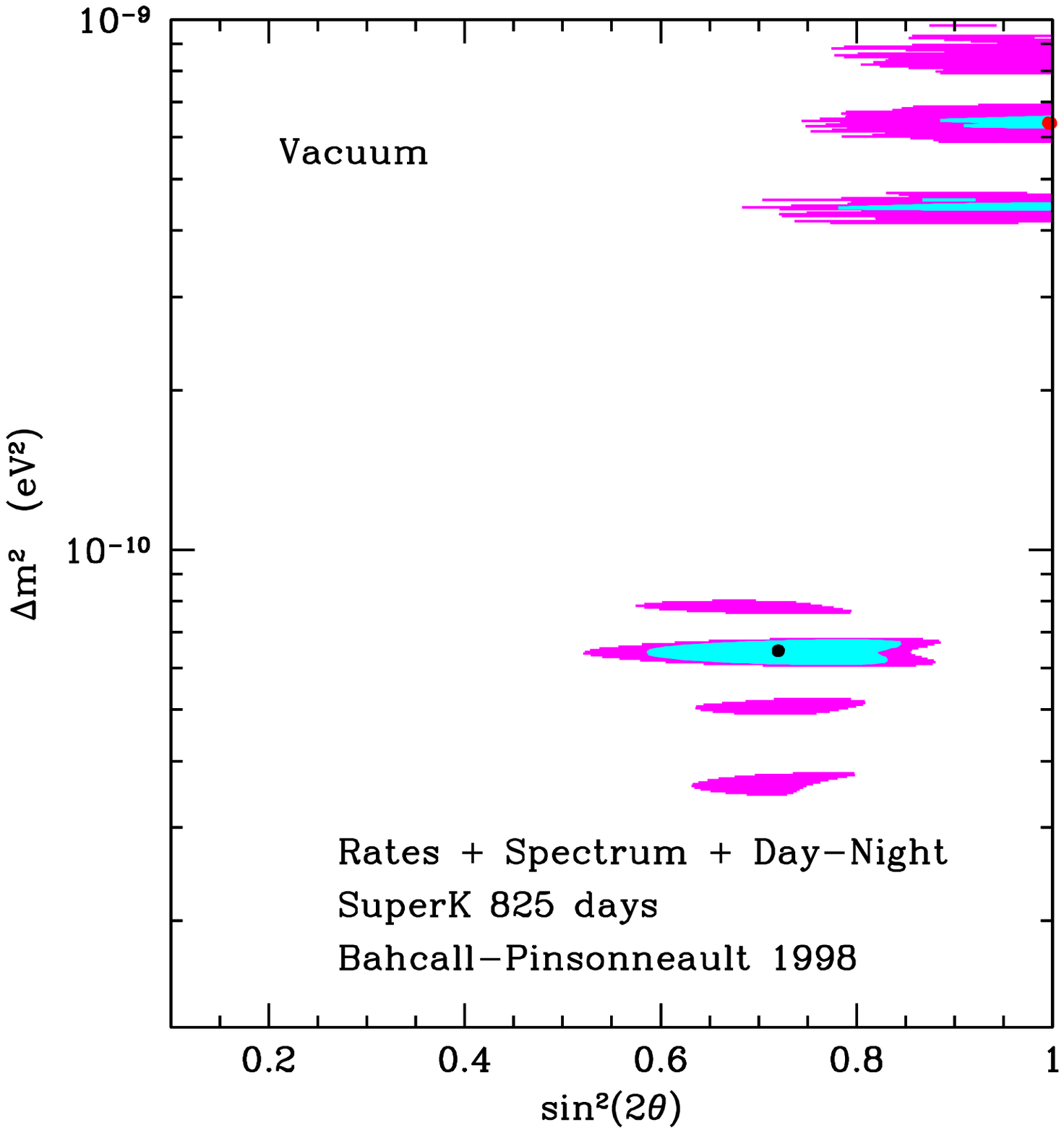,width=6cm}}
\vspace{10pt}
\hspace{3pt}
\caption[]{Regions of $\Delta m^2$ and $\sin^2 2 \theta$ for
$\nu-e \rightarrow \nu_x$ oscillations favoured by a global
analysis of solar neutrino data~\cite{bahcallpage}.}
\label{fig4}
\end{figure}
There is no hint what (combination of) other flavours the $\nu_e$ might be
oscillating into. In the atmospheric case, $\Delta m^2 \sim (2$ to $6)\times 10^{-3}$
eV$^2$ and large mixing are required. The Super-Kamiokande and Chooz data both exclude
$\nu_\mu\rightarrow\nu_e$ dominance, and zenith-angle distributions in the Super-Kamiokande
data favour $\nu_\mu\rightarrow\nu_\tau$ over $\nu_\mu$ oscillations into sterile neutrinos
$\nu_s$.

The past year has witnessed many theoretical studies of
neutrino masses~\cite{vast}, of which I now pick out just a few key
features: \\
\\
{\it Other light neutrinos?} We know from the LEP neutrino-counting 
constraint~\cite{LEPEWWG}, that any additional neutrinos
must be sterile $\nu_s$,
with no electroweak interactions or quantum numbers. 
But if so, what is to prevent
them from acquiring large masses: $M_s\nu_s\nu_s$ with $M_s \gg m_W$, as for the
$\nu_R$ discussed above? In the absence of some new theoretical superstructure,
this is an important objection to simply postulating light $\nu_s$ or $\nu_R$.\\
\\
{\it Majorana masses?} Most theorists expect the light neutrinos to be
essentially pure $\nu_L$, with only a small admixture ${\cal O}(m/M)$ of
$\nu_R$. In this
case, one expects the dominant effective neutrino mass 
term to be of Majorana type
$m_{eff}\nu_L\nu_L$, as given by (\ref{ten}) or (\ref{twelve}). \\
\\
{\it Large mixing?} Small neutrino mixing used perhaps to be favoured, by
analogy with the Cabibbo-Kobayashi-Maskawa mixing of quarks. However, theorists
now realize that this is by no means necessary. For one thing, the off-diagonal
entries in (now considered as a 3$\times$3 matrix) (\ref{twelve}) need not be
$\propto m_q$ or $m_\ell$~\cite{ELLN}. Moreover, even if $m\propto m_\ell$, we
have no independent
evidence that mixing is small in the lepton sector. Finally, even if $m$ were to
be approximately diagonal in the same flavour basis as 
the charged leptons $e, \mu, \tau$, 
why should this also be the same case for the  heavy Majorana matrix
$M$~\cite{ELLN}?\\
\\
{\it Could neutrinos be degenerate?} Are masses $\overline{m} \gappeq$ 2 eV and close
to the direct and astrophysical limits allowed \cite{EL}? Any such
scenario would need to respect the stringent constraint imposed by the absence of
$\beta\beta_{0\nu}$ decay~\cite{betabeta}:
\beq
<m_\nu>_e ~\simeq~ \overline{m} ~\vert c^2_{12} c^2_{13} e^{i \alpha} + s^2_{12}c^2_{13}
e^{i\beta} + s^2_{13}\vert\lappeq 0.2~{\rm eV}
\label{thirteen}
\eeq
In view of the upper limit on $\nu_\mu - \nu_e$ mixing from the Chooz
experiment~\cite{Chooz},
let us neglect provisionally the last term in (\ref{thirteen}). In this case,
there must be a cancellation between the first two terms, requiring
$\alpha\simeq\beta + \pi$, and
\beq
c^2_{12}-s^2_{12} = \cos 2\theta_{12} \lappeq 0.1 \Rightarrow \sin^2 2\theta_{12}
\gappeq 0.99
\label{fourteen}
\eeq
Thus maximal $\nu_e-\nu_\mu$ mixing is necessary if the neutrinos are
heavy and degenerate. This certainly excludes the
small-mixing-angle MSW solution and possibly even the large-mixing-angle MSW
solution, since this is not compatible with $\sin^2 2\theta = 1$ (which would
yield a constant energy-independent suppression of the solar neutrino flux), and
global fits typically indicate that $\sin^2 \theta_{12} \lappeq$ 0.97, as 
seen in Fig.~5~\cite{BKS}.
\begin{figure}[htb] 
\centerline{\epsfig{file=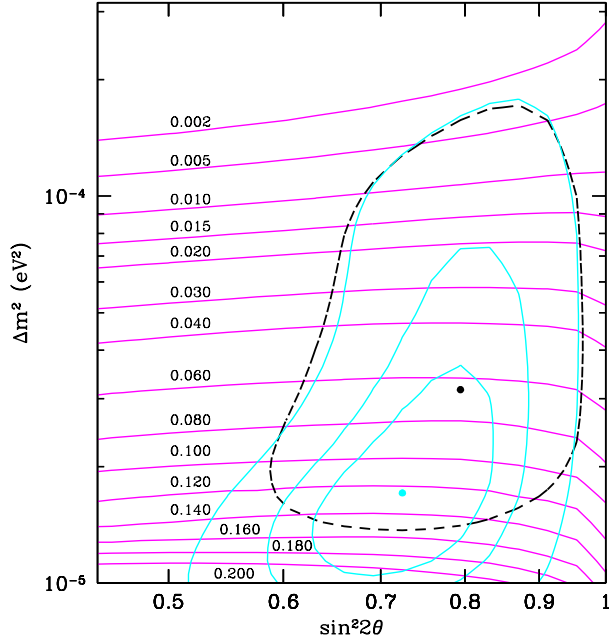,width=8cm}}
\vspace{10pt}
\hspace{3pt}
\caption[]{Preferred region of $\sin^2 \theta$ and $\Delta m^2$
for the large-mixing-angle MSW solution to the solar neutrino
problem, both with (dashed contours) and without (grey contours)
the measured day-night asymmetry: note that $\sin^2 \theta <
0.97$~\cite{BKS}.}
\label{fig5}
\end{figure}
 Global fits before the new
Super-Kamiokande data on the energy
spectrum indicated that $\sin^2 2\theta \sim 1$ was possible for
vacuum-oscillation solutions. However, the new Super-Kamiokande analysis of the
energy spectrum now indicates~\cite{Totsuka} that, if there is any
consistent vacuum-oscillation
solution at all, it may
well have $\sin^2 2\theta$ considerably below 1,
providing
another potential nail in the coffin of degenerate neutrinos.

The vacuum-oscillation solution would require, moreover,  extreme degeneracy:
$\Delta m \sim 10^{-10} \overline{m}$, which is impossible to reconcile with a
simple calculation of neutrino mass renormalization in models with degenerate
masses at the $m_{\nu_R}$ scale~\cite{EL}.
Mass-renormalization effects
also endanger the
large-angle MSW solution (which would require $\Delta m \sim 10^{-4}
\overline{m}$), and, in the context of bimaximal mixing models, also generate
unacceptable values of the neutrino mixing angles.
These renormalization problems may not be insurmountable~\cite{Casas},
but
they do raise
non-trivial issues that must be addressed in 
models~\cite{Barbieri} of (near-)degenerate neutrino masses.
Our provisional conclusion is that $m_{\nu_i} \ll$ 2 eV, 
with the most likely case being
$m_{\nu_i} \lappeq \sqrt{\Delta m^2_{atmo}} \sim$ 0.06 eV, 
too small to be of much interest
for hot dark matter.

\section{The Lightest Supersymmetric Particle}
 
 The motivation for supersymmetry at an accessible energy is provided by 
the gauge
 hierarchy problem~\cite{hierarchy}, namely that of understanding why $m_W
\ll m_P$, the 
only
 candidate for a fundamental mass scale in physics. Alternatively and
 equivalently, one may ask why $G_F\sim g^2/m^2_W \gg G_N = 1/m^2_P$,
where $M_P$ is the Planck mass, expected to be the fundamental
gravitational mass scale. Or 
one may
 ask why
 the Coulomb potential inside an atom is so much
 larger than the Newton potential, which is equivalent to why $e^2 = 
{\cal O}(1) \gg
 m_pm_e/m^2_P$, where $m_{p,e}$ are the proton and electron masses.

One might think it would be sufficient to choose the bare mass
 parameters: $m_W\ll m_P$. However, one must then contend with quantum
 corrections, which are quadratically divergent:
 \beq
 \delta m^2_{H,W} = {\cal O}~\left({\alpha\over\pi}\right)~\Lambda^2
 \label{fifteen}
 \eeq
 These are much larger than $m_W$, if the cutoff $\Lambda$ representing 
the
 appearance of new physics is taken to be ${\cal O}(m_P)$. This means 
that one must
 fine-tune the bare mass parameter so that it is almost exactly cancelled 
by the
 quantum correction (\ref{fifteen}) in order to obtain a small physical 
value of
 $m_W$. This seems unnatural, and the alternative is to introduce new 
physics at
 the TeV scale, so that the correction (\ref{fifteen}) is naturally 
small.
 
 At one stage, it was proposed that this new physics might correspond to 
the Higgs
 boson being composite~\cite{technicolour}. However, calculable scenarios
of this type are
 inconsistent with the precision electroweak data from LEP and elsewhere. 
The
 alternative is to postulate approximate supersymmetry~\cite{susy}, whose
pairs of 
bosons and
 fermions produce naturally cancelling quantum corrections:
 \beq
 \delta m^2_W = {\cal O}~\left({\alpha\over\pi}\right)~\vert m^2_B - 
m^2_F\vert
 \label{sixteen}
 \eeq
 that are naturally small: 
 $\delta m^2_W \lappeq m^2_W$ if 
\beq
\vert m^2_B - m^2_F\vert \lappeq \; \; {\rm 1 TeV}^2.
\label{seventeen}
\eeq
 There are many other possible motivations for supersymmetry, some of which 
are discussed
below, but this is  the only
 one that gives reason to expect that it might be accessible to the 
current
 generation of accelerators and in the range  expected for a 
cold
 dark matter particle.
 
 The minimal supersymmetric extension of the Standard Model
(MSSM)~\cite{MSSM} has the same
 gauge interactions as the Standard Model, and the Yukawa interactions 
are very
 similar:
 \beq
 \lambda_d QD^cH + \lambda_\ell LE^cH + \lambda_u QU^c\bar H + \mu\bar HH
 \label{eighteen}
 \eeq
 where the capital letters denote supermultiplets with the same quantum 
numbers as
 the left-handed fermions of the Standard Model. The couplings
 $\lambda_{d,\ell,u}$ give masses to down quarks, leptons and up quarks
 respectively, via distinct Higgs fields $H$ and $\bar H$, which are 
required in
 order to cancel triangle anomalies. The new parameter in 
(\ref{eighteen}) is the
 bilinear coupling $\mu$ between these Higgs fields, that plays a 
significant
 r\^ole in the description of the lightest supersymmetric particle, as we 
see
 below. The gauge quantum numbers do not forbid the appearance of 
additional
 couplings~\cite{Dreiner}
 \beq
 \lambda LLE^c + \lambda^\prime LQD^c + \lambda U^cD^cD^c
 \label{nineteen}
 \eeq
 but these violate lepton or baryon number, and we assume
they are 
absent.
 One significant aspect of the MSSM is that the quartic scalar 
interactions are
 determined, leading to important constraints on the Higgs mass, as we 
also see
 below.
 
 Supersymmetry must be broken, since  supersymmetric partner particles do 
not have
 identical masses, and this is usually parametrized by scalar mass 
parameters
 $m^2_{0_i}\vert\phi_i\vert^2$, gaugino masses ${1\over 2} M_a\tilde
 V_a\cdot\tilde V_a$ and trilinear scalar couplings $A_{ijk}\lambda_{ijk}
 \phi_i\phi_j\phi_k$. These are commonly supposed to be inputs from some
 high-energy physics such as supergravity or string theory. It is often
 hypothesized that these inputs are universal: $m_{0_i} \equiv m_0, 
M_a\equiv
 m_{1/2}, A_{ijk}\equiv A$, but these assumptions are not strongly 
motivated by any
 fundamental theory. The physical sparticle mass parameters are then 
renormalized  in a calculable way:
 \beq
 m^2_{0_i} = m^2_0 + C_i m^2_{1/2}~,~~ M_a = \left({\alpha_a\over
 \alpha_{GUT}}\right)~~m_{1/2}
 \label{twenty}
 \eeq
where the $C_i$ are calculable coefficients~\cite{renorm}
 and MSSM phenomenology is then parametrized by $\mu, m_0, m_{1/2}, A$ 
and
 $\tan\beta$ (the ratio of Higgs v.e.v.'s).
 
 Precision electroweak data from LEP and elsewhere provide two 
qualitative
 indications in favour of supersymmetry. One is that the inferred 
magnitude of
 quantum corrections favour a relatively light Higgs boson~\cite{LEPEWWG}
 \beq
 m_h \lappeq 200~{\rm GeV}
 \label{twentyone}
 \eeq
 which is highly consistent with the value predicted in the MSSM: $m_h 
\lappeq$
 150 GeV~\cite{susymh} as a result of the constrained quartic couplings.
(On the other 
hand,
 composite Higgs models predicted an effective Higgs mass $\gappeq$ 1 TeV
and other unseen quantum corrections.)  The other indication in favour 
of
 low-energy supersymmetry is provided by measurements of the gauge 
couplings at
 LEP, that correspond to $\sin^2 \theta_W \simeq 0.231$ in agreement with 
the
 predictions of supersymmetric GUTs with sparticles weighing about 1~TeV, 
but
 in disagreement with non-supersymmetric GUTs that predict
 $\sin^2\theta_W \sim 0.21$ to 0.22~\cite{sintheta}.  Neither of these
arguments provides 
an accurate
estimate of the sparticle mass scales, however, since they are both only 
logarithmically
sensitive to $m_0$ and/or $m_{1/2}$.

The lightest supersymmetric particle (LSP) is expected to be stable in 
the MSSM, and hence
should be present in the Universe today as a cosmological relic from the 
Big Bang~\cite{EHNOS,Keith}.  This is
a consequence of a multiplicatively-conserved quantum number called $R$ 
parity, which is
related to baryon number, lepton number and spin:
\beq
R = (-1)^{3B+L+2S}
\label{twentytwo}
\eeq
It is easy to check that $R = +1$ for all Standard Model particles and $R 
= -1$ for all
their supersymmetric partners.  The interactions (\ref{eighteen}) conserve
$R$, whilst those in  (\ref{nineteen}) would  violate $R$, in contrast to
 a Majorana neutrino mass term or the other interactions in minimal $SU(5)$ or 
$SO(10)$ GUTs. 
There are three important consequences of $R$ conservation: (i) 
sparticles are always
produced in pairs, e.g., $pp \to \tilde{q} \tilde{g} X$, $e^+ e^- \to 
\tilde{\mu}^+
\tilde{\mu}^-$,  (ii) heavier sparticles decay into lighter sparticles, 
e.g., $\tilde{q}
\to q \tilde{g}$,
$\tilde{\mu} \to \mu \tilde{\gamma}$, and (iii) the LSP is stable because 
it has no legal
decay mode.

If such a supersymmetric relic particle had either electric charge or 
strong interactions,
it would have condensed along with ordinary baryonic matter during the 
formation of
astrophysical structures, and should be present in the Universe today in 
anomalous heavy
isotopes.  These have not been seen in studies of $H$, $He$, $Be$, $Li$, 
$O$, $C$, $Na$,
$B$ and $F$ isotopes at levels ranging from $10^{-11}$ to
$10^{-29}$~\cite{Smith}, 
which are far below
the calculated relic abundances from the Big Bang:
\beq
\frac{n_{relic}}{n_p} \; \gappeq \; 10^{-6} \; \mbox{to} \; 10^{-10}
\label{twentythree}
\eeq
 for relics with electromagnetic or strong interactions. Except
possibly for very heavy relics, one would expect these primordial relic
particles to condense into galaxies, stars and planets, along with
ordinary bayonic material, and hence show up as an anaomalous heavy
isotope of one or more of the elements studied. There
would also be a `cosmic rain' of
such relics~\cite{Nussinov}, but this would presumably not be the dominant
source
of such particles on earth. The conflict with
(\ref{twentythree}) is sufficiently acute that
the lightest supersymmetric
relic must presumably be electromagnetically neutral and weakly
interacting~\cite{EHNOS}.  In particular, I
believe that the possibility of a stable gluino can be excluded.  
This leaves as
scandidates for cold dark matter a sneutrino $\tilde{\nu}$ with spin 0, 
some neutralino
mixture of $\tilde{\gamma} / \tilde{H}^0 / \tilde{Z}$ with spin 1/2, and 
the gravitino
$\tilde{G}$ with spin 3/2.

LEP searches for invisible $Z^0$ decays require $m_{\tilde{\nu}} \, 
\gappeq \, 43 \;
\mbox{GeV}$~\cite{EFOS}, and searches for the interactions of relic
particles with 
nuclei then enforce 
$m_{\tilde{\nu}} \, \gappeq \,$ few  TeV~\cite{Klap}, so we exclude this
possibility 
for the
LSP.  The possibility of a gravitino $\tilde{G}$ LSP has attracted 
renewed interest
recently with the revival of gauge-mediated models of supersymmetry 
breaking~\cite{GR}, and could
constitute warm dark matter if $m_{\tilde{G}} \simeq 1 \, \mbox{keV}$.  
In this talk,
however, I concentrate on the $\tilde{\gamma} / \tilde{H}^0 /
\tilde{Z}^0$ neutralino combination $\chi$, which is the best 
supersymmetric candidate for
cold dark matter.

The neutralinos and charginos may be characterized 
at the tree level by three parameters: 
$m_{1/2}$, $\mu$
and tan$\beta$.  The lightest neutralino $\chi$ simplifies in the limit 
$m_{1/2} \to 0$
where it becomes essentially a pure photino $\tilde{\gamma}$, or $\mu \to 
0$ where it
becomes essentially a pure higgsino $\tilde{H}$.  These possibilities are 
excluded,
however, by LEP and the FNAL Tevatron collider~\cite{EFOS}.  From the
point of view 
of astrophysics and
cosmology, it is encouraging that there are generic domains of the 
remaining parameter
space where $\Omega_{\chi}h^2 \simeq 0.1$ to $1$, in particular in 
regions where $\chi$ is
approximately a $U(1)$ gaugino $\tilde{B}$, as seen in
Fig.~6~\cite{EFGOS}.
\begin{figure}[htb] 
\centerline{\epsfig{file=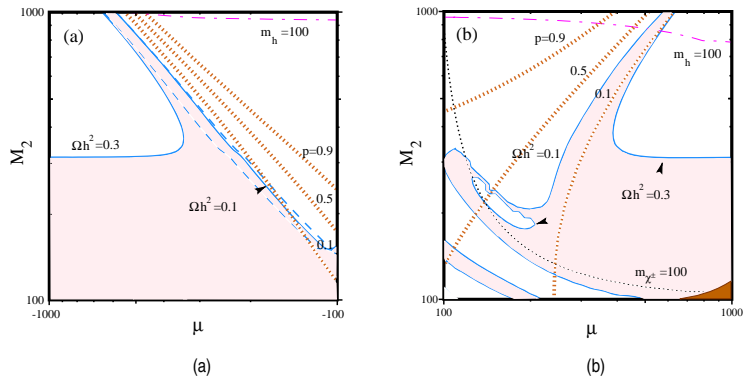,width=10cm}}
\vspace{10pt}
\hspace{3pt}
\caption[]{Regions of the $(\mu, M_2)$ plane in which the
supersymmetric relic density may lie within the interesting range
$0.1 \le \Omega h^2 \le 0.3$~[14].}
\label{fig6}
\end{figure}

Purely experimental searches at LEP enforce $m_{\chi} \gappeq 30$ GeV \cite{LEPC}. 
This bound can be strengthened by making various theoretical assumptions, 
such as the
universality of scalar masses $m_{0_i}$, including in the Higgs sector, 
the cosmological
dark matter requirement that $\Omega_{\chi} h^2 \leq 0.3$ and the 
astrophysical preference
that $\Omega_{\chi} h^2 \geq 0.1$.  Taken together as in Fig.~7,
\begin{figure}[htb] 
\centerline{\epsfig{file=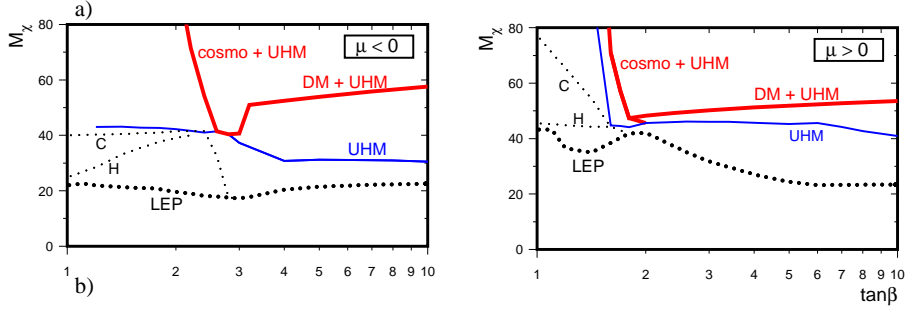,width=12cm}}
\vspace{12pt}
\hspace{3pt}
\caption[]{Theoretical lower limits on the lightest neutralino
mass, obtained by using the unsuccessful Higgs searches (H), the
cosmological upper limit on the relic density (C), the assumption that
all input scalar masses are universal, including those of the Higgs
multiplets (UHM), and combining this with the cosmological upper (cosmo)
and astrophysical lower (DM) limits on the cold dark matter
density~\cite{EFOS}.}
\label{fig7}
\end{figure}
 we see 
that they enforce
\beq
m_{\chi} \gappeq 50 \; \mbox{GeV}
\label{twentyfour}
\eeq
Moreover, LEP has already explored almost all the parameter 
space available for a
Higgsino-like LSP, and this possibility will also be thoroughly explored 
by the full running of LEP~\cite{LEPC}.

Should one be concerned that no sparticles have yet been seen by either 
LEP or the FNAL
Tevatron collider?  One way to quantify this is via the amount of 
fine-tuning of the input
parameters required to obtain the physical value of $m_W$~\cite{fine}:
\beq
\Delta_o = 
Max_{i} \;
\mid \frac{a_i}{m_W} \; \frac{\partial m_W}{\partial a_i} \mid
\label{twentyfive}
\eeq
where $a_i$ is a generic supergravity input parameter.  
The LEP
exclusions impose~\cite{CEOP}
\beq
\Delta_o \gappeq 8
\label{twentysix}
\eeq
Although fine-tuning is a matter of taste, this is perhaps not large 
enough to be alarming,
and could in any case be reduced significantly if a suitable theoretical 
relation between
some input parameters is postulated~\cite{CEOP}. Moreover, 
it is interesting to
note that the 
amount of
fine-tuning $\Delta_o$ is minimized when $\Omega_{\chi}h^2 \sim 0.1$ as
preferred astrophysically, as seen in Fig. 8~\cite{CEOPO}. 
\begin{figure}[htb] 
\centerline{\epsfig{file=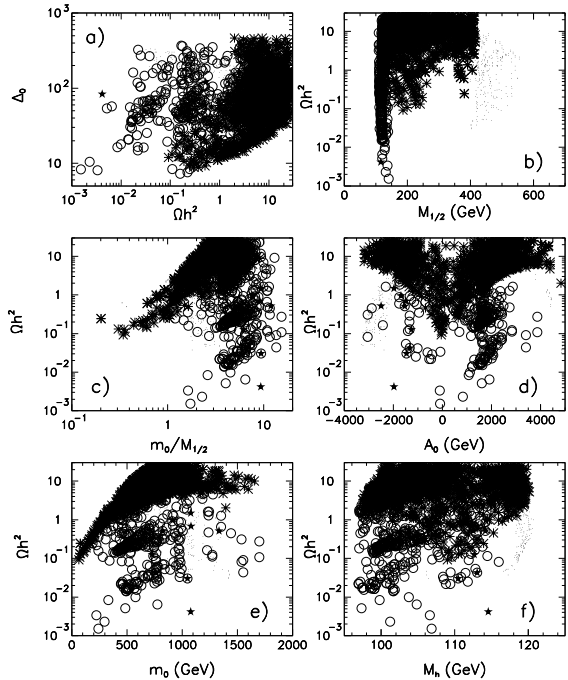,width=8cm}}
\vspace{10pt}
\hspace{3pt}
\caption[]{The correlation~\cite{CEOP} between the fine-tuning price
$\Delta_0$
and the relic density $\Omega h^2$, showing dependences on model
parameters.}
\label{fig8}
\end{figure}
This means that
solving the 
gauge hierarchy
problem naturally leads to a relic neutralino density in the range of 
interest to
astrophysics and cosmology.  I am unaware of any analogous argument for 
other particle dark matter candidates such as the neutrino or the
axion.

For certain ranges of the MSSM parameters, our present electroweak vacuum is
unstable against the development of vev's for $\tilde q$ and $\tilde l$
fields, leading to vacua that would break charge and colour conservation.
Among the dangerous possibilities are flat directions of the effective
potential in which combinations such as $L_i Q_3 D_3,~~H_2L_i,~~LLE,~~ H_2L$
acquire vev's. Avoiding these vacua imposes constraints that depend on the
soft supersymmetry breaking parameters: they are weakest for $A\simeq
m_{1/2}$. Fig. 9 illustrates some of the resulting constraints in the
$(m_{1/2}, m_0)$ plane, for different values of $\tan\beta$ and signs of
$\mu$ \cite{AF}. We see that they cut out large parts of the plane, particularly for low
$m_0$. In combination with cosmology, they tend to rule out large values of
$m_{1/2}$, but this aspect needs to be considered in conjunction with the
effects of coannihilation, that are discussed in the next paragraph.

As $m_{\chi}$ increases, the LSP annihilation cross section decreases and 
hence its relic
number and mass density increase. How heavy could the LSP be?  Until 
recently, the limit
given was $m_{\chi} \lappeq 300$ GeV~\cite{limit}.  However, it has now
been pointed 
out that there are
regions of the MSSM parameter space where coannihilations of the $\chi$ 
with the stau
slepton $\tilde{\tau}$ could be important, as also seen in Fig. 9~\cite{EFO}.
\begin{figure}[htb] 
\centerline{\epsfig{file=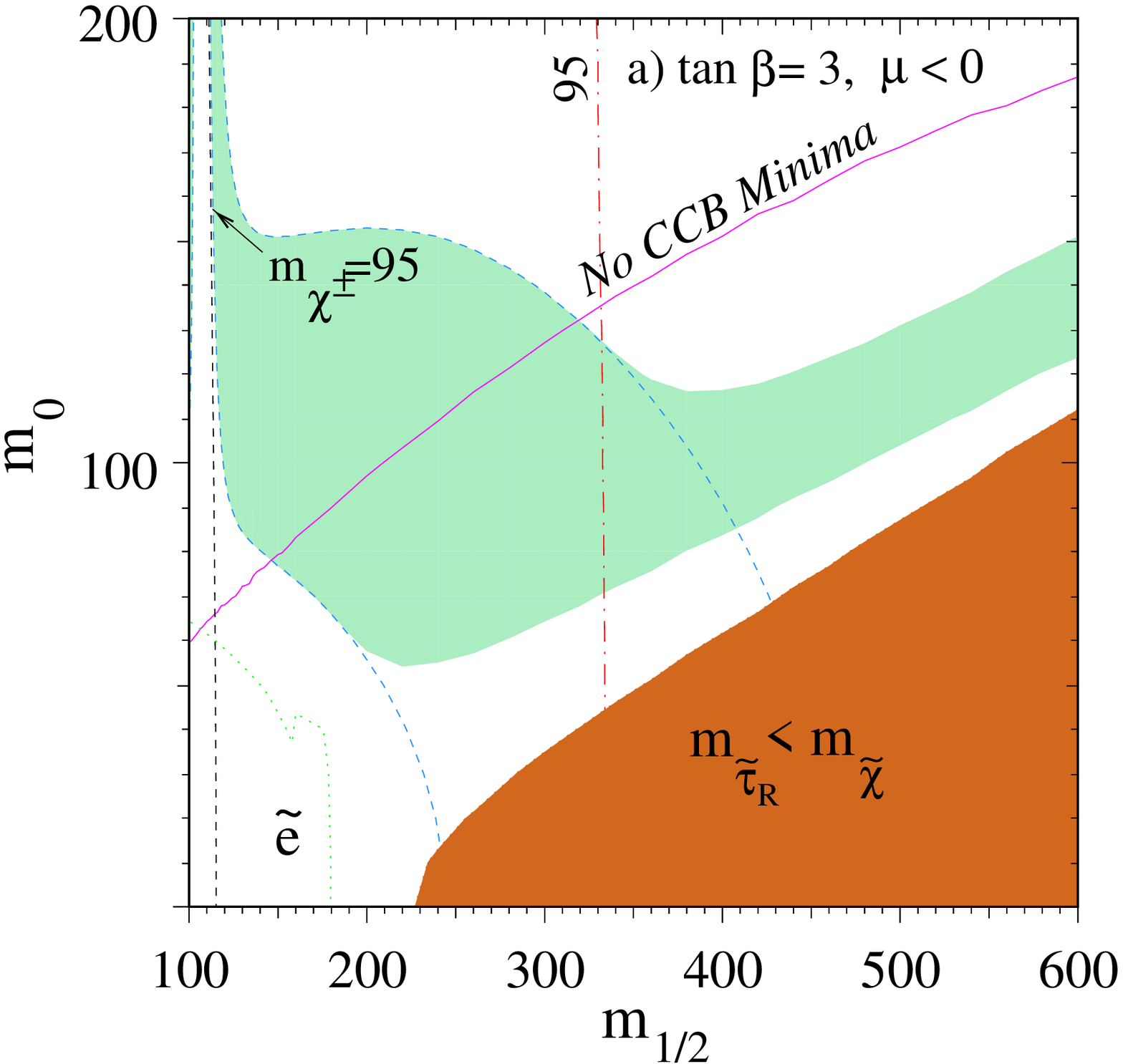,width=4cm}
\epsfig{file=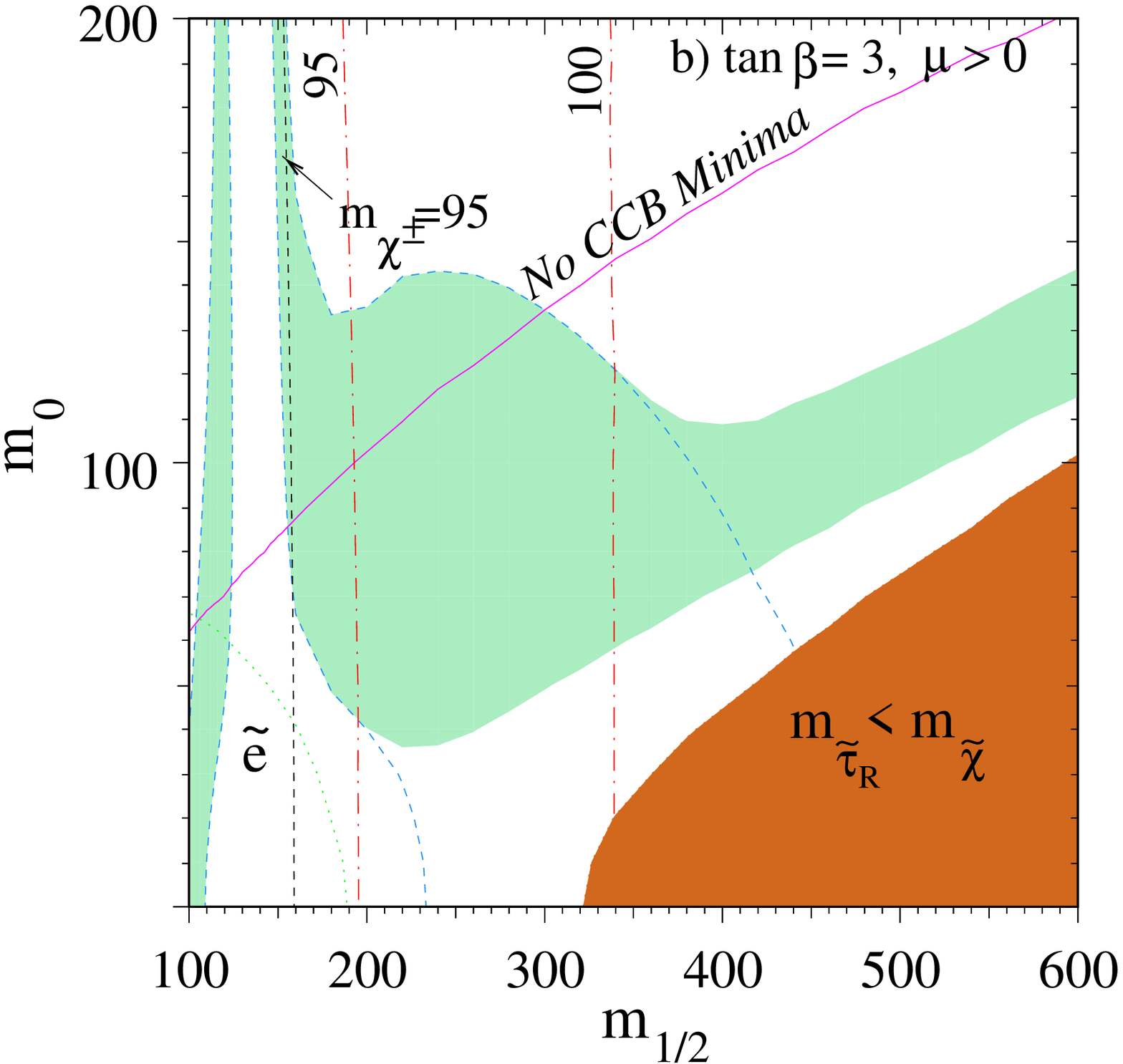,width=4cm}}
\centerline{\epsfig{file=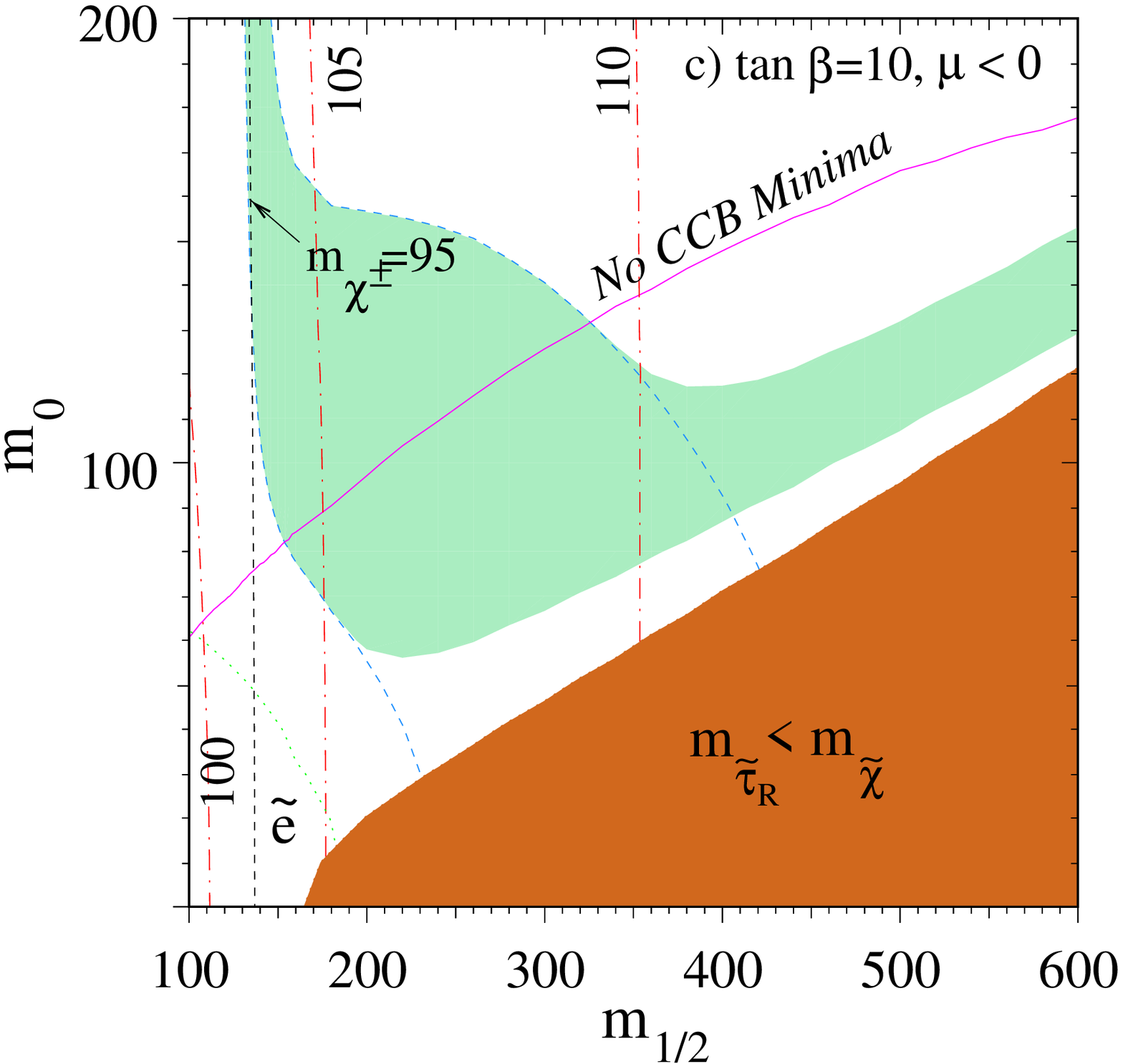,width=4cm}
\epsfig{file=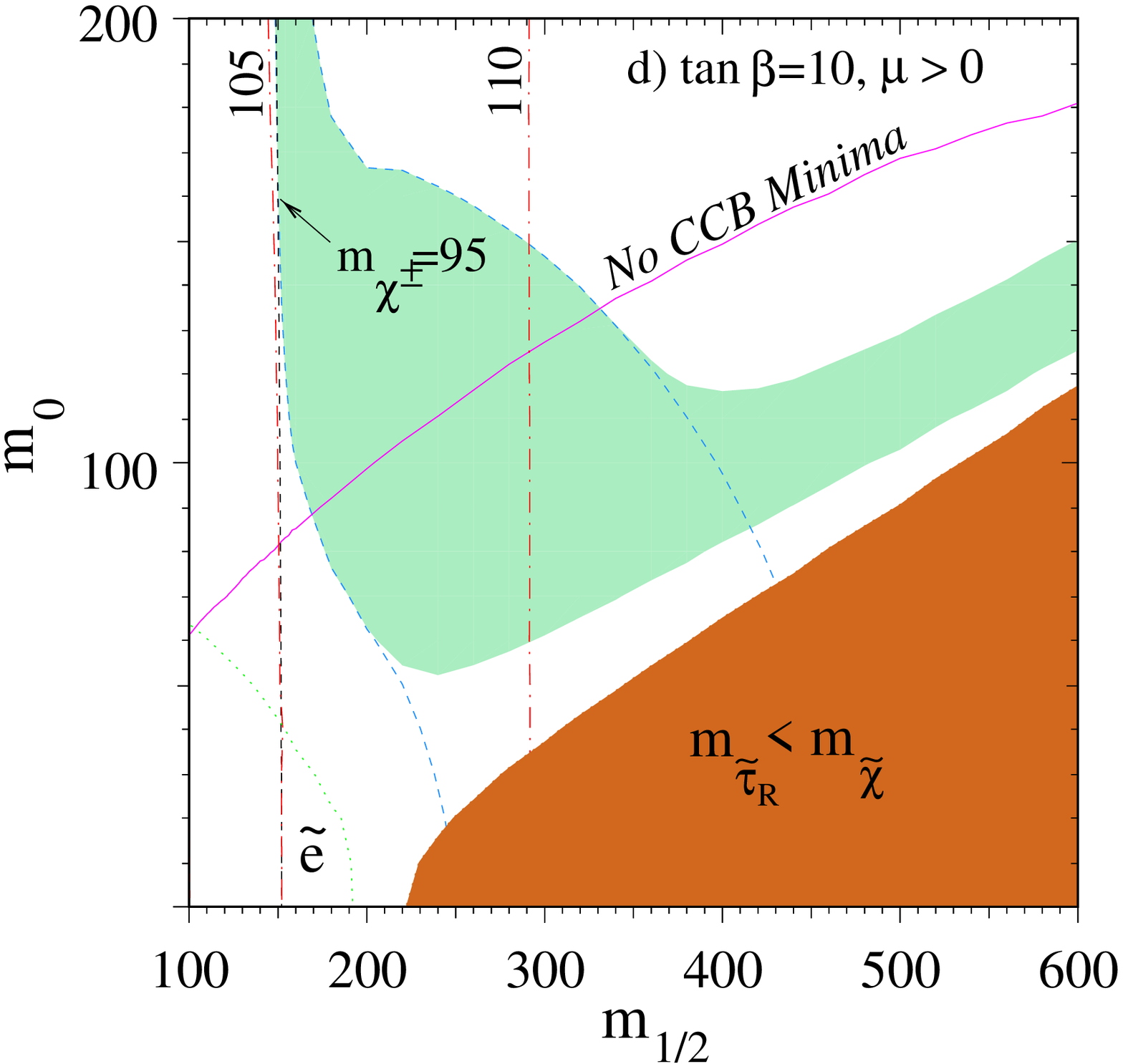,width=4cm}}
\vspace{10pt}
\hspace{3pt}
\caption[]{The light shaded region is that favoured by calculations
of the relic density of LSPs, including coannihilation effects, which
are significant on the right sides of the panels~\cite{EFO}. The dark
shaded
region is excluded because it would have charged dark matter. Also
indicated are mass contours of interest to LEP searches, and a
potential lower bound on $m_0$ obtained by requiring that the
true vacuum not break charge and colour conservation (CCB)~\cite{AF}.}
\label{fig9}
\end{figure}
These 
coannihilations would
suppress $\Omega_{\chi}$, allowing a heavier neutralino mass, and we now 
find that~\cite{EFO}
\beq
m_{\chi} \lappeq \; 600 \, \mbox{GeV}
\label{twentyseven}
\eeq
is possible if we require $\Omega_\chi h^2 \leq 0.3$.  In the past, it was thought that all the 
cosmologically-preferred region of
MSSM parameter space could
be explored by the LHC~\cite{Abdullin}, as seen in Fig. 10, 
but it now seems
possible that there may be a delicate region close to the upper bound 
(\ref{twentyseven}). 
This point requires further study.
\begin{figure}[htb] 
\centerline{\epsfig{file=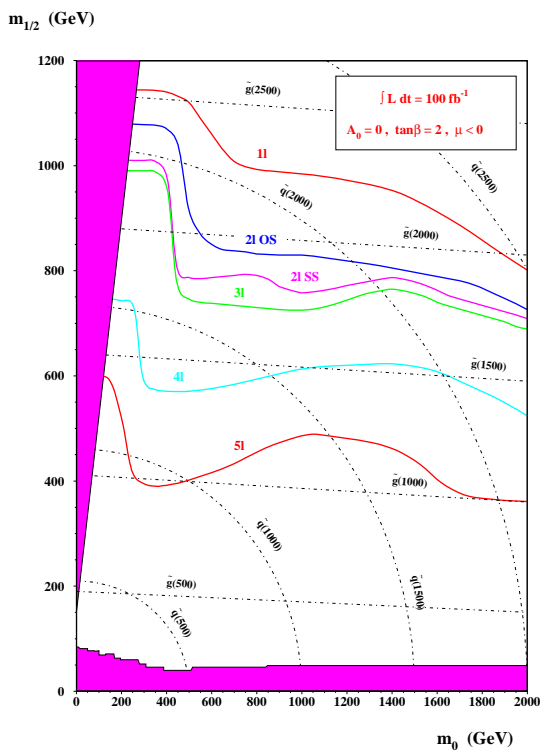,width=8cm}}
\vspace{10pt}
\hspace{3pt}
\caption[]{The region of the $(m_0, m_{1/2})$ plane accessible
to sparticle searches at the LHC [22].  }
\label{fig10}
\end{figure}

 \section{Searches for Dark Matter Particles}
\subsection{Annihilation in the Galactic Halo}

One strategy to look for dark matter particles is via their annihilations in
the galactic halo; $\chi\chi\rightarrow\ell^+\ell^-, \bar q q\rightarrow \bar
p, e^+, \gamma, \nu$ in the cosmic rays~\cite{halo}. Figure 11
\begin{figure}[htb] 
\centerline{\epsfig{file=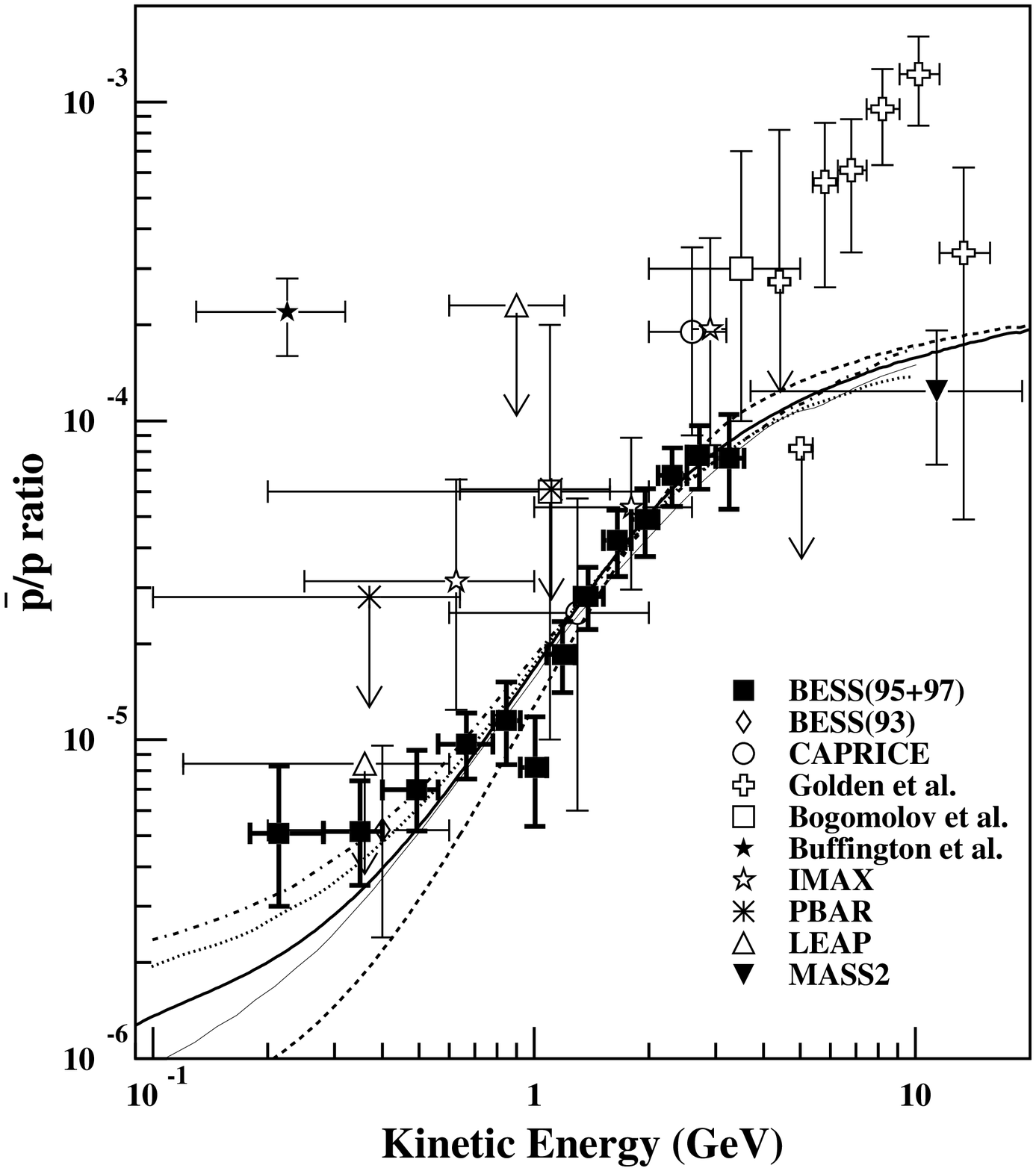,width=6cm}}
\vspace{10pt}
\hspace{3pt}
\caption[]{Data on $\bar p$'s in the cosmic rays. The recent BESS
data~\cite{BESS}
agree with calculations based on secondary $\bar p$ production by
primary matter cosmic rays.}
\label{fig4}
\end{figure}
 shows the current
measurements of cosmic-ray $\bar p$'s. The lines indicate the secondary flux
expected to be produced by primary matter cosmic rays. Some of the earlier
measurements were above this conventional expectation, fuelling speculation
about possible exotic sources such as $\chi\chi$ annihilation. However, the
recent BESS data~\cite{BESS} agree very well with conventional secondary
production. There
may still be some scope for exotic sources at low energies $E\lappeq$ 300 MeV
or at higher energies $E\gappeq$ 3 GeV, and one of the objectives of the AMS
experiment~\cite{AMS} is to explore this possibility. Figure 12 
\begin{figure}[htb] 
\centerline{\epsfig{file=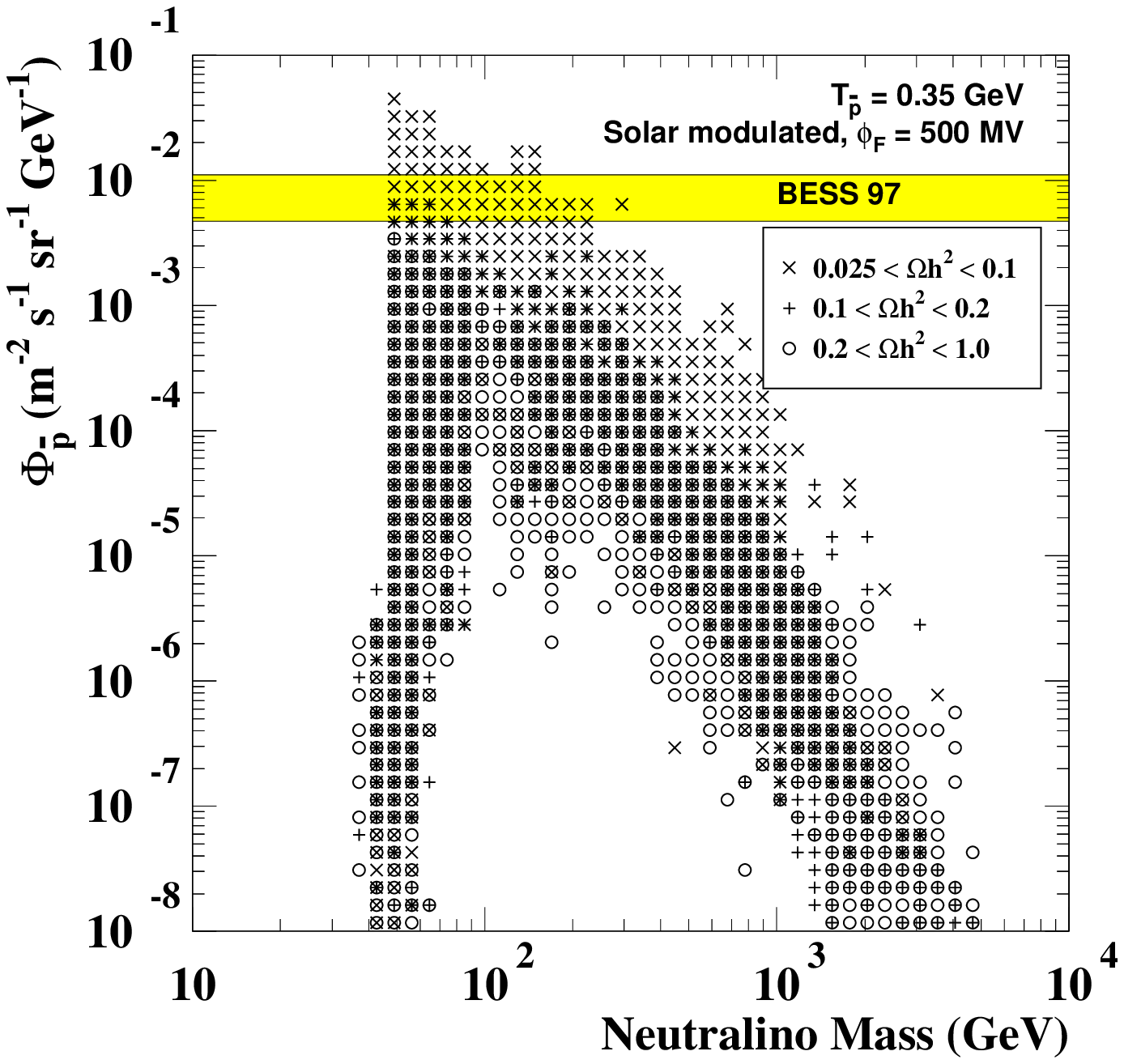,width=8cm}}
\vspace{10pt}
\hspace{3pt}
\caption[]{The fluxes of cosmic-ray $\bar p$'s predicted in a sampling of
superymmetric models~\cite{Bergpbar} confronted with the BESS
data~\cite{BESS}.}
\label{fig12}
\end{figure}
shows that some
supersymmetric models are already excluded by the BESS data~\cite{BESS},
and indicates how big an opportunity AMS may have.

It has been suggested~\cite{HEAT} that there may be an excess of
cosmic-ray positrons at
energies above  1 GeV, as seen in Fig. 13,
\begin{figure}[htb] 
\centerline{\epsfig{file=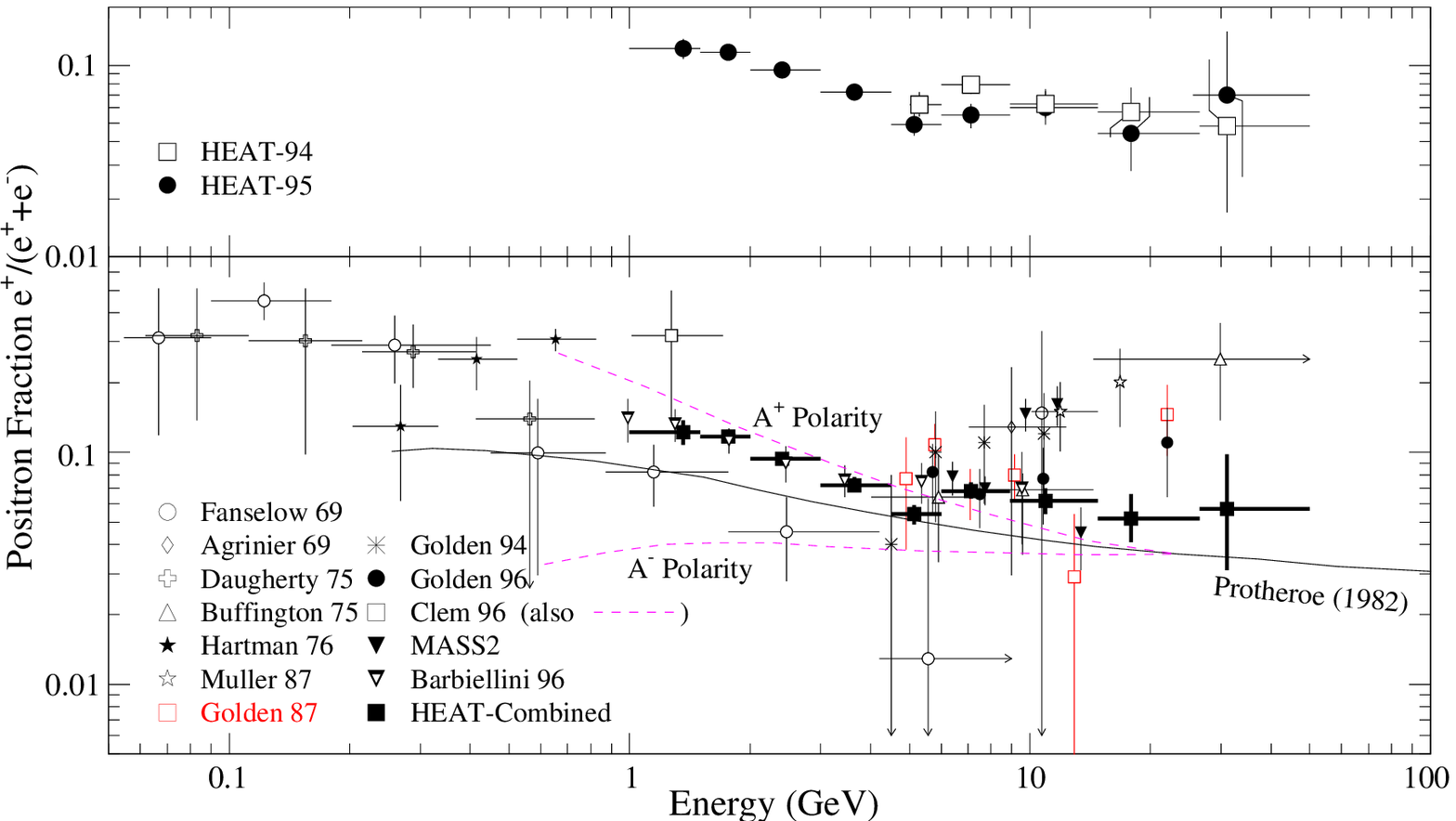,width=8cm}}
\vspace{10pt}
\hspace{3pt}
\caption[]{Data on the fraction of positrons in the cosmic
rays~\cite{HEAT}, compared with standard leaky-box model calculations.}
\label{fig13}
\end{figure}
although the uncertainties in the standard leaky-box
model prevent any definite conclusion at this stage. AMS has reported an excess
of positrons at lower energies: $E~\lappeq$~1~GeV, but it seems unlikely that
these have an exotic origin. It may in the future be able to contribute to the
clarification of the possible excess at higher energies.

Data from EGRET on $\gamma$ rays above 100 MeV have been
interpreted~\cite{EGRET} as
containing an excess from directions centred around the galactic centre. These
could well be due to some unresolved astrophysical sources or some diffuse
mechanism such as inverse Compton scattering. If due to dark matter
annihilation, they would require $<\sigma v> \sim (10^{-24}$ to $10^{-25}$)
cm$^2$ if the dark matter  particles are not clumped. This range is above
that allowed for supersymmetric dark matter: $<\sigma v> \sim 3\times
10^{-27}/(\Omega h^2)$ cm$^2$. Therefore it has been
suggested~\cite{Berggamma} that the dark
matter may be clumped, as in some models of structure formation. A
phenomenological approach to this possibility is to calculate the maximal
clumpiness enhancement allowed for any supersymmetric model, taking into
account the experimental upper limits on, e.g., $\bar p$ annihilation products,
and then calculate the maximum possible $\gamma$ flux. Some representative
calculations~\cite{Berggamma} are shown in Fig. 14: 
\begin{figure}[htb] 
\centerline{\epsfig{file=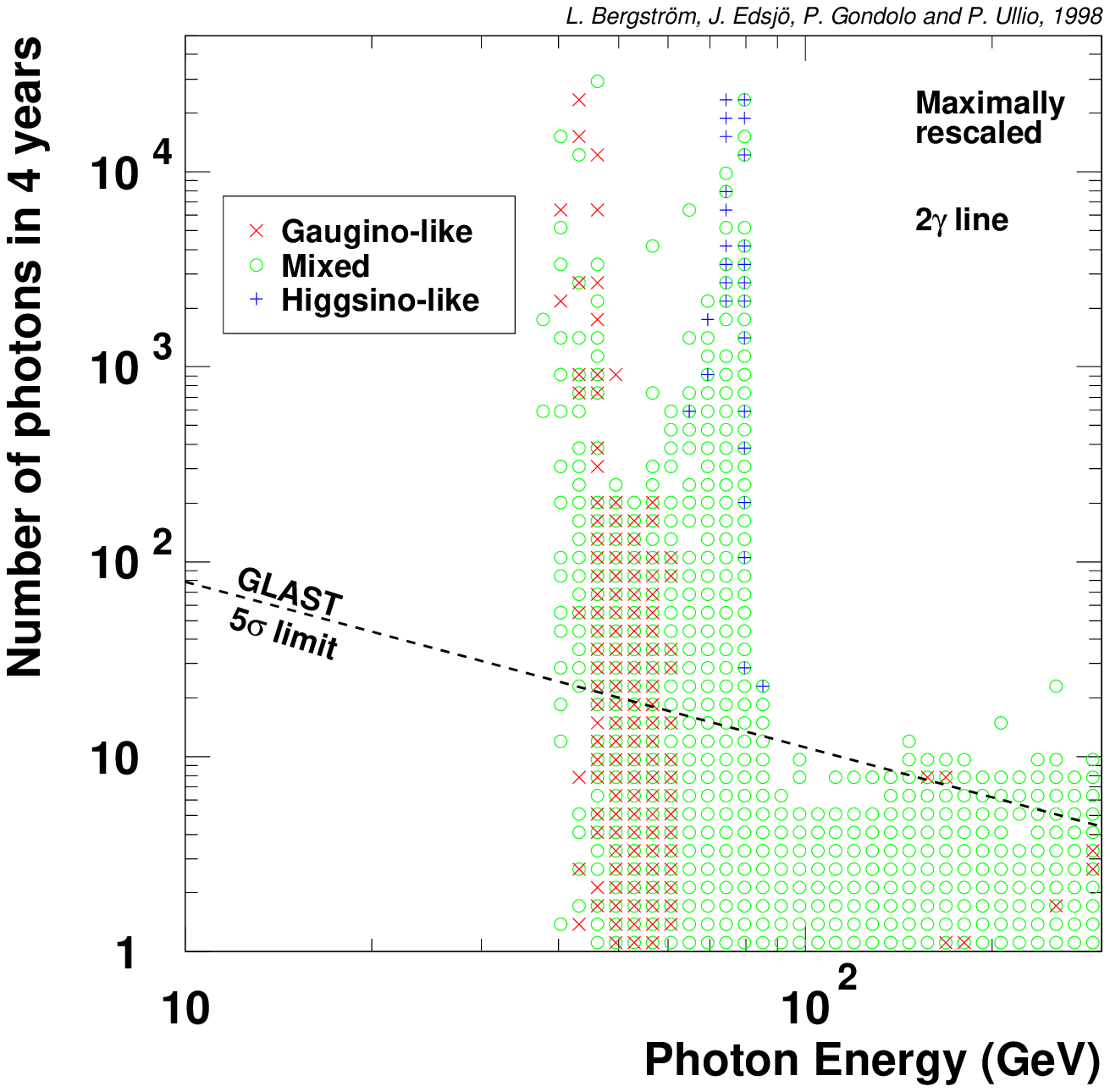,width=6cm}
\epsfig{file=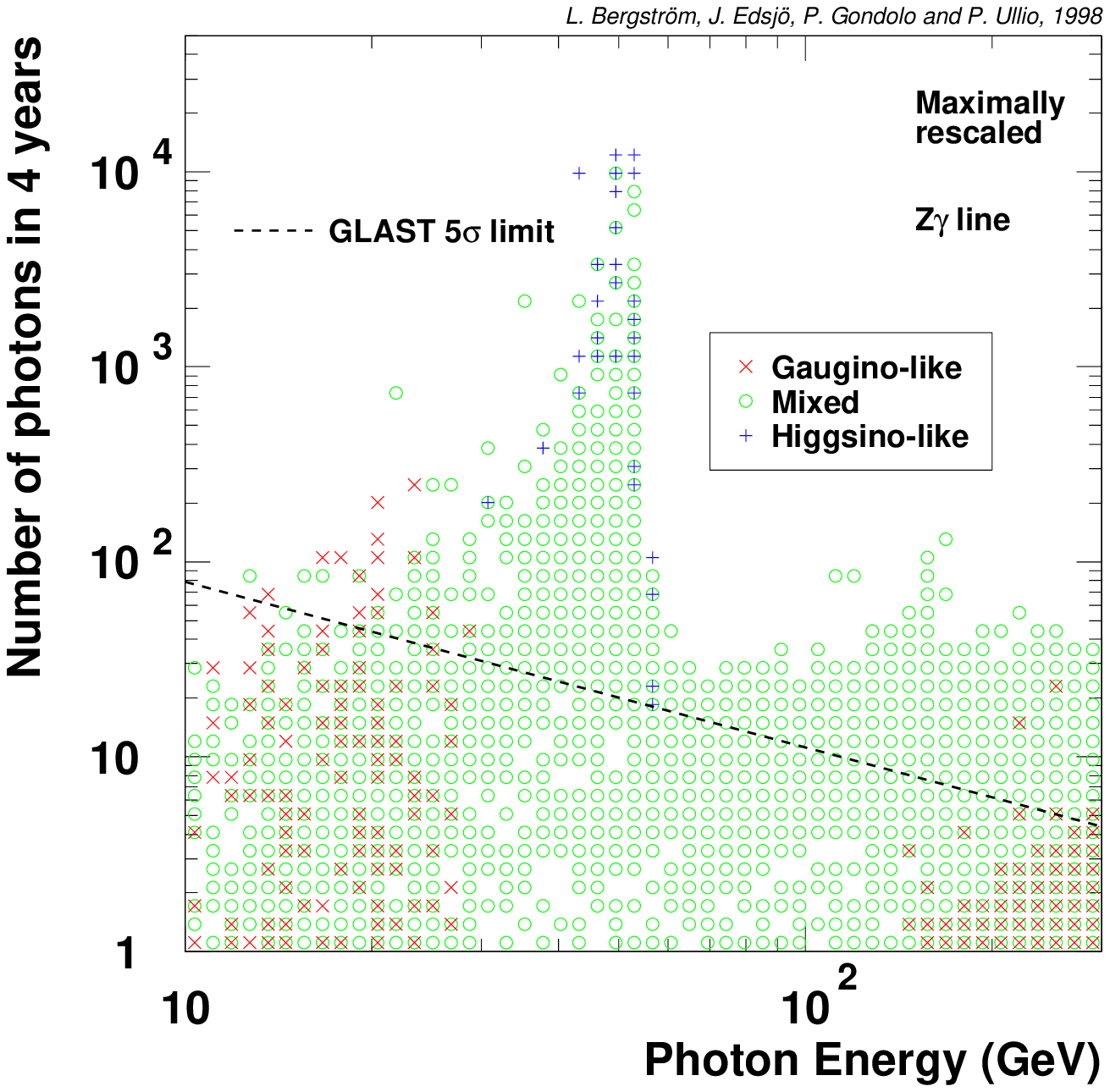,width=6cm}}
\vspace{10pt}
\hspace{3pt}
\caption[]{Possible fluxes of cosmic-ray $\gamma$'s allowed if
one postulates the maximal clumping of cold dark matter in the galactic
halo~\cite{Berggamma}, compared with the expected GLAST sensitivity.}
\label{fig14}
\end{figure}
we see that the $\gamma$ fluxes could in
principle be orders of magnitude larger than those detectable by GLAST.

\subsection{Annihilation in the Sun or Earth}

A dark matter passing through the Sun (Earth) may scatter on some nucleus
inside, losing source recoil energy which may convert its orbit from hyperbolic
to elliptical, with perhelion (perigee) below the surface. Then it will scatter
repeatedly, eventually settling into a quasi-thermal distribution beneath the
surface. This population of relic particles is controlled by annihilation:
$\chi\chi\rightarrow\ell^+\ell^-, \bar qq$, yielding as observable products
energetic neutrinos: $E_\nu~\gappeq$~1~GeV. These may be detected either
directly in an underground detector, or indirectly via $\mu$'s produced in
material surrounding the detector.

As seen in Fig. 15~\cite{Bergnu}, 
\begin{figure}[htb] 
\centerline{\epsfig{file=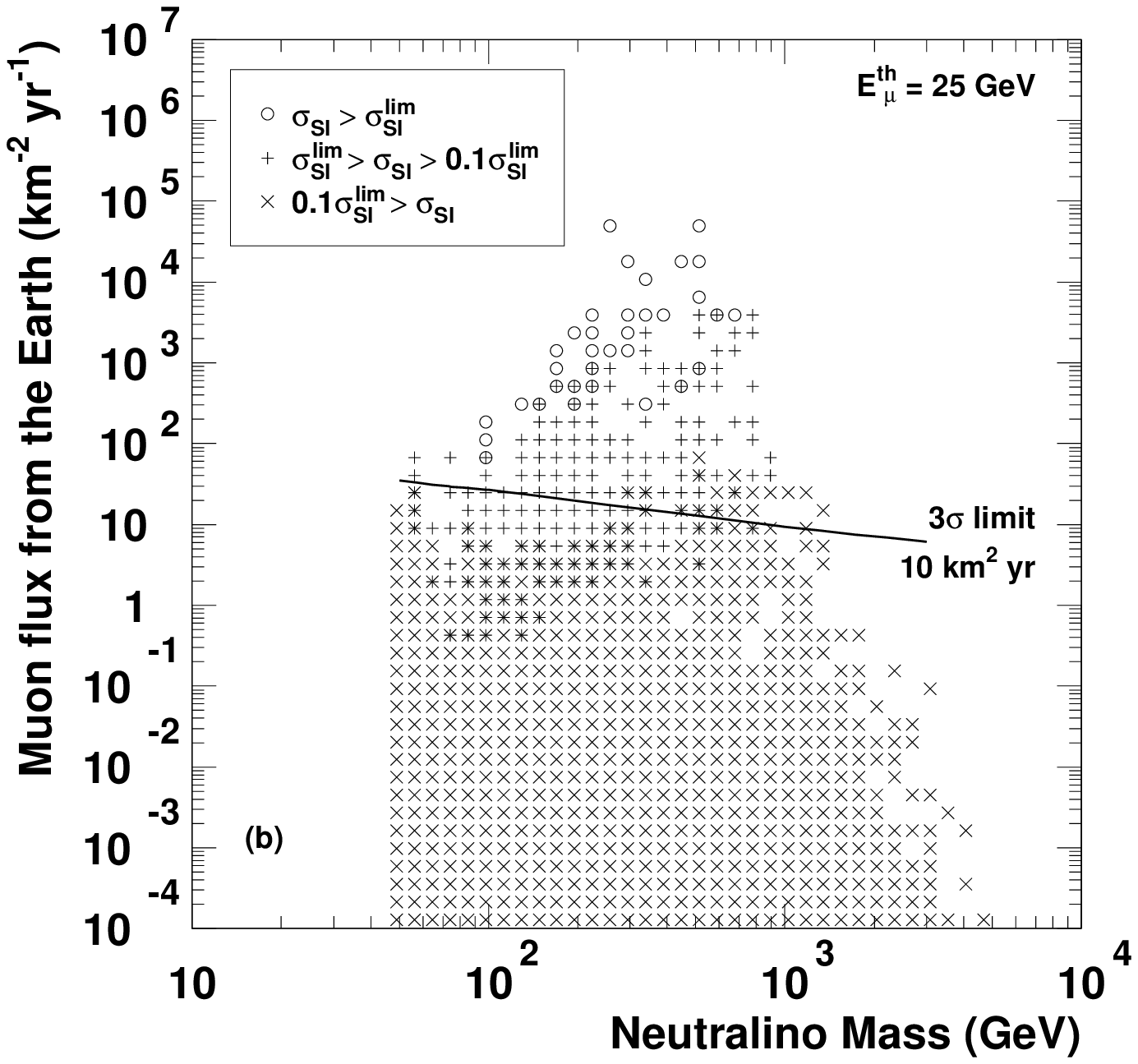,width=6cm}
\epsfig{file=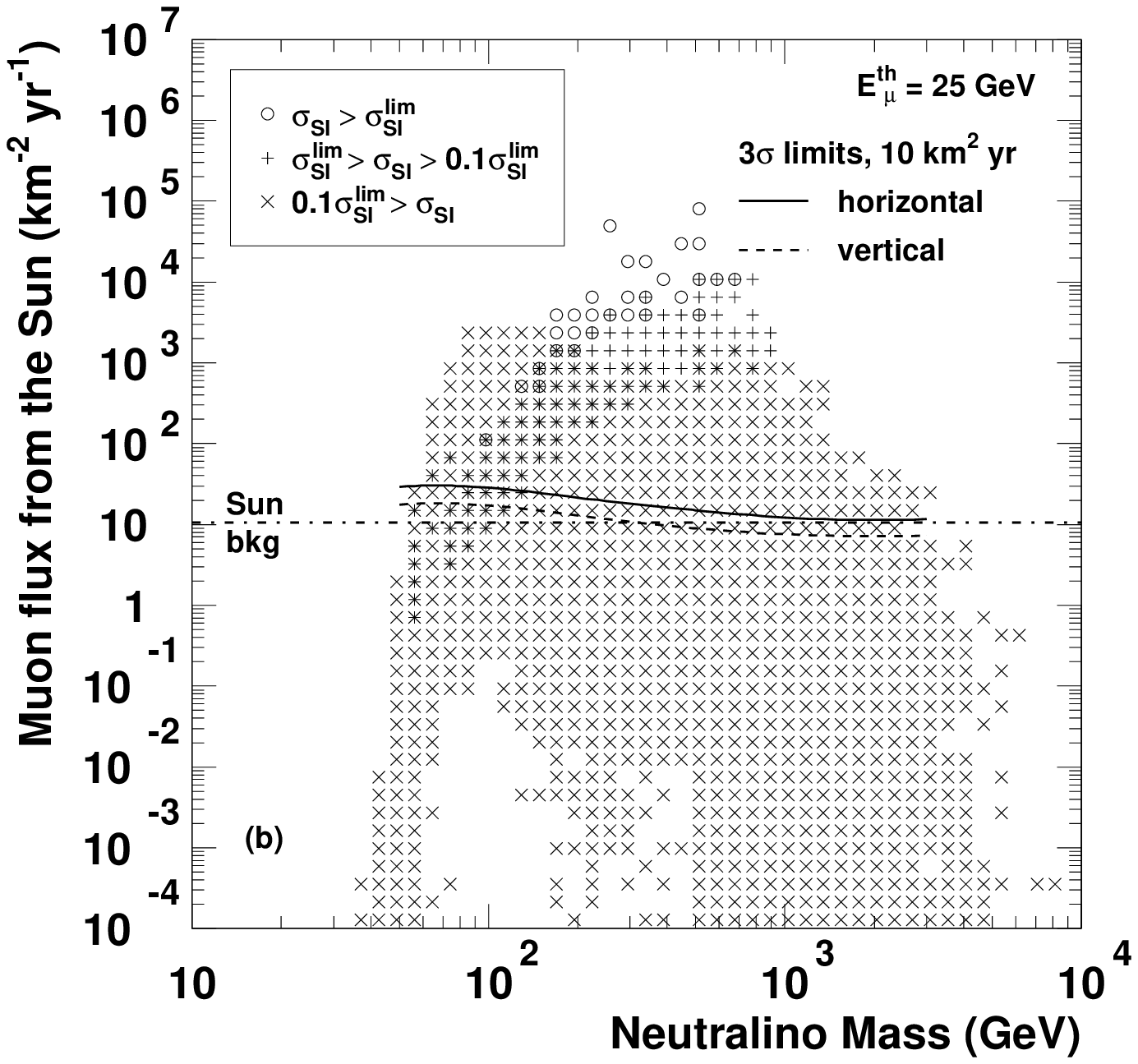,width=6cm}}
\vspace{10pt}
\hspace{3pt}
\caption[]{The fluxes of high-energy $\mu$'s due to the interactions
of $\nu$'s produced by relic annihilations inside the Sun or Earth,
as produced by a sampling of supersymmetric models~\cite{Bergnu},
compared with the expected sensitivity of a 1~km$^2$ detector.}
\label{fig15}
\end{figure}
a 1 km$^2$ muon detector would be able to detect quite a
number of supersymmetric models that do not produce detectable cosmic-ray $\bar
p$ fluxes, via either the solar or subterranean cosmic rays they produce. It
has recently been pointed out~\cite{Damour} that there could be an
enhancement of relic
annihilations in the Earth due to a solar-system population of relic particles
that is augmented by Jupiter's gravitational field, so the prospects may be
even brighter than indicated in Fig. 15.

\subsection{Direct Detection of Dark Matter}

Many experiments around the world are looking for the recoil energy deposited
in low-background underground detectors by relic scattering on nuclei. The
typical recoil energy deposited is ${\cal O}(m_\chi v^2/2) \sim$ tens of keV.
The interaction may be mediated by squark, $Z$ and Higgs exchanges, which
contribute both spin-dependent and spin-independent matrix elements. The former
are related to the quark contributions to the proton spin $(\Delta q)$ and the
latter to the quark contributions to the proton mass. The time-dependent
contributions are important for some light nuclei such as Fluorine, where they
can be calculated quite reliably. However, the spin-independent
contributions
are coherent and more important for heavier nuclei~\cite{GJK}.

Figure 16 
\begin{figure}[htb] 
\centerline{\epsfig{file=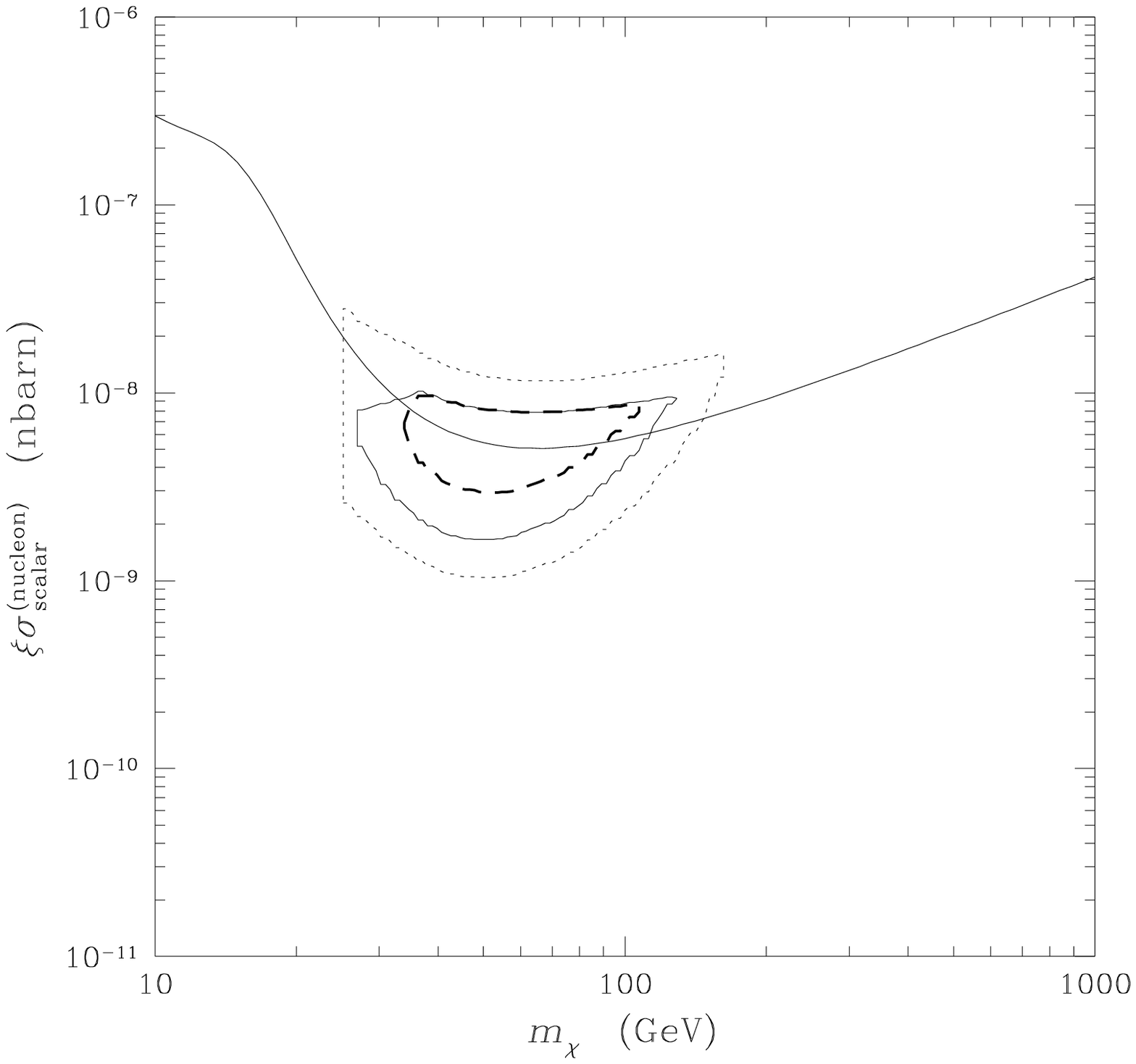,width=6cm}}
\vspace{10pt}
\hspace{3pt}
\caption[]{Upper limit (solid line) and regions (dotted, solid 
and dashed lines) not excluded by the DAMA~\cite{DAMA} search for an
annual modulation effect.}
\label{fig16}
\end{figure}
shows the upper limit for scattering on an individual nucleon,
together with the region of parameter space that could not be excluded by an
experiment searching for an annual modulation of the detector rate in a
particular range of deposited energy~\cite{DAMA}. Although on the large
side, the range of
cross section in this non-excluded domain is not inconsistent with some model
calculations. Future experiments will be able to improve the current upper
limits shown in Fig. 16 by several orders of magnitude and explore much of the
model parameter space.

What if there were a real signal in the non-excluded region in Fig. 16? It
would correspond to a supersymmetric relic weighing $\sim$ 50 to 100 GeV. Such 
models might well produce an observable cosmic-ray $\bar p$ flux or $\mu$ flux from
high-energy solar or subterranean neutrinos~\cite{Bottino}.

\subsection{Supersymmetry at the LHC}

These searches for astrophysical sparticles must compete with accelerator
searches. LEP has almost completed the exploration of its available kinematic
reach, and the Tevatron has a window for possible sparticle discoveries.
However, the best prospects for the discovery of supersymmetry will be offered
by the LHC. It will benefit from large cross sections for squark $(\tilde q)$
and gluino $(\tilde g)$ production, and their cascade decays into lighter
sparticles offer many opportunities for distinctive signatures. As seen in Fig.
10~\cite{Abdullin}, it should be
possible to detect $m_{\tilde q/\tilde g} \lappeq$ 2 to 2.5 TeV, and some
detailed spectroscopic measurements will be possible. The entire supersymmetric
dark matter region were $\Omega_\chi h^2 \lappeq$ 0.3 will be covered by the
LHC. So those looking for supersymmetric dark matter via cosmic rays or other
astrophysical signatures should hurry up, and do their best before LHC startup
in 2005!

\section{Superheavy Relic Particles}

It has been suggested~\cite{Dim} that cold dark matter particles should
weigh $\lappeq$ 1
TeV, as exemplified by the MSSM range 
(\ref{twentyseven}). This expectation is based
on the assumption that the cold dark matter particles were at one time in 
thermal
equilibrium.  However, much heavier relic
particles are 
possible if one
invokes non-thermal production mechanisms.  Non-thermal decays of 
inflatons in conventional
models of cosmological inflation could yield $\Omega_{\chi} \sim 1$ for 
$m_{\chi} \sim
10^{13}$ GeV.  Preheating via the parametric resonance decay of the 
inflaton could
even yield 
$\Omega_{\chi} \sim 1$ for $m_{\chi} \sim 10^{15}$ GeV.  Other 
possibilities include a
first-order phase transition at the end of inflation, and gravitational 
relic production
induced by the rapid change in the scale factor in the early
Universe~\cite{Zilla}.  
It is therefore of
interest to look for possible experimental signatures of superheavy dark 
matter.

One such possibility is offered by ultra-high-energy cosmic rays.  Those 
coming from
distant parts of the Universe $(D \gappeq 100 Mpc)$ are expected to be 
cut off at an energy
$E \lappeq 5 \times 10^{19}$ GeV, because of the reaction $p + 
\gamma_{CMBR} \to \Delta^+$~\cite{GZK}.
However, as discussed extensively here, no such Greisen-Zatsepin-Kuzmin cut-off
is seen in the data \cite{UHECR} The
ultra-high-energy cosmic rays must originate nearby, and 
(unless the intergalactic magnetic field is unexpectedly high~\cite{Biermann})
should point 
back to any
point-like sources such as AGNs~\cite{MT}.  However, no such discrete
sources have been 
identified as yet.

Could the ultra-high-energy cosmic rays be due to the decays of 
superheavy relic particles?
These should be clustered in galactic haloes (including our own), and 
hence give an
anisotropic flux~\cite{anisotropy}, but there would be no obvious point
sources. There 
have been some
reports of anisotropies in high-energy cosmic rays, but it is not clear 
whether they could
originate in superheavy relic decays.

We  analyzed~\cite{BEN}  possible superheavy relic candidates
in string~\cite{ELN}
and/or $M$ theory.
One expects Kaluza-Klein states when six excess dimensions are 
compactified: $10 \to 4$ or
$11 \to 5$, which we call {\it hexons}.  However, these are expected to 
weigh $\gappeq
10^{16}$ GeV, which may be too heavy, and there is no particular reason 
to expect hexons to
be metastable. In $M$ theory, one expects massive states associated with 
a further
compactification: $5 \to 4$ dimensions, which we call {\it pentons}.  
Their mass could be
$\sim 10^{13}$ GeV, which would be suitable, but there is again no good 
reason to expect
them to be metastable.  We are left with bound states from the hidden 
sector of string/$M$
theory, which we call {\it cryptons}~\cite{ELN}.  These could also have
masses $\sim 
10^{13}$ GeV, and
might be metastable for much the same reason as the proton in a GUT, 
decaying via
higher-dimensional multiparticle operators.  For example, in a flipped 
$SU(5)$ model we
have a hidden-sector $SU(4) \times SO(10)$ gauge group, and the former 
factor confines
four-constituent states which we call {\it tetrons}.  Initial
studies~\cite{ELN,BEN} indicate that the
lightest of these might well have a lifetime $\gappeq 10^{17} y$, which 
would be suitable
for the decays of superheavy dark matter particles.
Detailed simulations
have been made of the spectra of particles produced 
by the
fragmentation of their decay products~\cite{Berez,BS}, and the
ultra-high-energy 
cosmic-ray data are
consistent with the decays of superheavy relics weighing $\sim 10^{12}$ 
GeV, as seen in
Fig. 17~\cite{BS}.  
\begin{figure}[htb] 
\centerline{\epsfig{file=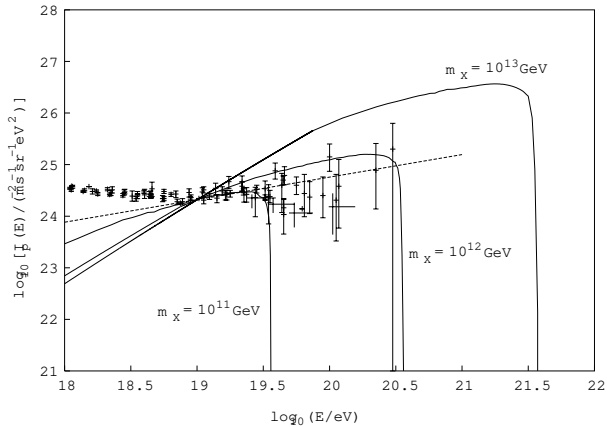,width=8cm}}
\vspace{10pt}
\hspace{3pt}
\caption[]{The ultra-high energy cosmic ray flux compared
with a model calculation based on the decays of superheavy relic
particles~\cite{BS}. }
\label{fig17}
\end{figure}
 Issues to be resolved here include the roles of
supersymmetric 
particles in the
fragmentation cascades, and the relative fluxes of $\gamma, \nu$ and $p$ 
among the
ultra-high-energy cosmic rays.

\section{Vacuum Energy}

As mentioned in Section 1, data on large-scale structure~\cite{Bahcall} and
high-redshift supernovae~\cite{highz} have recently 
converged on the
suggestion that the energy of the vacuum may be non-zero, as seen in 
Figs. 1, 3.
  In my view, 
this represents a
wonderful opportunity for theoretical physics: a number to be calculated 
in the Theory of
Everything including quantum gravity.  The possibility that the vacuum 
energy may be
non-zero may even appear more natural than a zero value, since there is 
no obvious symmetry
or other reason known why it should vanish.

In the above paragraph, I have used the term {\it vacuum energy} rather 
than {\it
cosmological constant}, because it may not actually be constant.  This 
option has been
termed {\it quintessence} in~\cite{Steinhardt}, which
discusses a 
classical scalar-field
model that is not strongly motivated by the Standard Model, supersymmetry 
or GUTs, though
something similar might emerge from string theory.  I prefer to think 
that a varying vacuum
energy might emerge from a quantum theory of gravity, as the
vacuum 
relaxes towards an
asymptotical value (zero?) in an infinitely large and old Universe. We 
have recently given~\cite{EMN}
an example of one such possible effect which yields a contribution to the 
vacuum energy
that decreases as $1/t^2$.  This is compatible with the high-redshift 
supernova data, and
one may hope that these could eventually discriminate between such a 
possibility and a true
cosmological constant.

\end{document}